\documentclass{aastex63}
\usepackage{amsmath}
\usepackage{natbib}
\usepackage{multirow}
\usepackage{graphicx}
\usepackage{amssymb}
\usepackage{subfigure}
\usepackage{lineno}

\hypersetup{linkcolor=magenta, citecolor=cyan, filecolor=yellow, urlcolor=blue}
\shorttitle{Power-Law Spectral Components in SGRBs}
\shortauthors{Tang et al}

\begin{document}

\title{Prevalence of Extra Power-Law Spectral Components in Short Gamma-Ray Bursts}
\author{Qing-Wen Tang}
\affiliation{Department of Physics, School of Science, Nanchang University, Nanchang 330031, China}
\author{Kai Wang}
\affiliation{Department of Astronomy, School of Physics, Huazhong University of Science and Technology, Wuhan 430074, China}
\author{Liang Li}
\affiliation{ICRANet, Piazza della Repubblica 10, I-65122 Pescara, Italy}
\author{Ruo-Yu Liu}
\affiliation{School of Astronomy and Space Science, Nanjing University, Xianlin Road 163, Nanjing 210023, China}
\affiliation{Key Laboratory of Modern Astronomy and Astrophysics (Nanjing University), Ministry of Education, Nanjing 210023, China}
\email{$\ast$ qwtang@ncu.edu.cn; kaiwang@hust.edu.cn; ryliu@nju.edu.cn}

\begin{abstract}
Prompt extra power-law (PL) spectral component is discovered in some bright gamma-ray bursts (GRBs), which usually dominates the spectral energy distribution below tens of keV or above $\sim$10 MeV. However, its origin is still unclear. In this paper, we present a systematic analysis of 13 \textit{Fermi} short GRBs as of August 2020,  with the contemporaneous keV--MeV and GeV detections during the prompt emission phase. We find that the extra PL component is a ubiquitous spectral feature for short GRBs, showing up in all 13 analyzed GRBs. The PL indices are mostly harder than $-$2.0, which may be well reproduced by considering the electromagnetic cascade induced by ultra-relativistic protons or electrons accelerated in the prompt emission phase. The average flux of these extra PL components positively correlates with that of the main spectral components, which implies they may share the same physical origin.
\end{abstract}


\section{Introduction}
\label{sec:Introduction}
Gamma-ray bursts (GRBs) are the most energetic explosions in the universe. They can be divided into two phenomenological categories based on their duration in the prompt phase, namely, long GRBs (LGRBs) and short GRBs (SGRBs), separated at about two seconds. Various physical models have been proposed to explain the prompt emission, such as the photospheric model \citep{Rees05, Giannios07, Peer08, Beloborodov11}, the internal shock model \citep{rees94, Kobayashi97, Daigne98} and the magnetic reconnection model \citep{Spruit01, ZhangB11}.
The spectral analysis is thus the key clue to investigate GRB radiation mechanism, and thus can help us to understand their underlying physical process. Observationally, GRB prompt emission exhibits diverse spectral properties. Those spectra in the keV--MeV energy range can generally be fitted by some empirical functions, such as the Band function (BAND component)~\citep{1993ApJ...413..281B}, the simple power law function (PL component), the PL with a high-energy exponential cutoff function (CPL component) and the smoothly broken PL function (SBPL component), based on the 10-year observations by the Gamma-ray Burst Monitor (GBM) onboard \textit{Fermi}~\citep{2021ApJ...913...60P}, hereafter the GBM catalog.

Combining the observations of the Large Area Telescope (LAT) onboard \textit{Fermi}, it is interesting to note that the keV--GeV spectra of some GRBs consist of more than one component. For example, a BAND component with a PL component for GRB 080916C, GRB 090510, GRB 0909026A and GRB 110731A~\citep{2010ApJ...709L.172R, 2010ApJ...716.1178A,2013ApJS..209...11A}, a CPL component with a PL component for GRB 090902B (time-integrated), GRB 100414A and GRB 160709A~\citep{2013ApJS..209...11A,2019ApJ...876...76T}, and a blackbody component (BB or multi-BB) with a PL component for GRB 081221, GRB 090902B (time-resolved), 110920A, GRB 160107A and GRB 160709A~\citep{2013ApJ...768..187B,2010ApJ...709L.172R,2015MNRAS.450.1651I,2018PASJ...70....6K,2019ApJ...876...76T}.

There are 186 GRBs reported in the 10-year catalog (hereafter LAT catalog) of \textit{Fermi}$-$LAT \citep{2019ApJ...878...52A}, from August 2008 to August 2018, among which 169 are LGRBs and 17 are SGRBs. In previous studies, only two SGRBs, namely, GRB~090510 and GRB~160709A were discovered to show the extra PL component in the spectrum~\citep{2010ApJ...716.1178A,2019ApJ...876...76T}. In order to further search for and explore the properties of the extra PL component, here we perform a comprehensive joint spectral analysis of \textit{Fermi}$-$GBM and \textit{Fermi}$-$LAT data of the selected short GRBs in the 8~keV--10~GeV energy range detected between August 2008 and August 2020. The same analysis for LGRBs will be performed and reported elsewhere.

The physical origin of the extra PL spectral components has been extensively explored in the framework of both the internal dissipation models \citep{2009ApJ...705L.191A,2009A&A...498..677B,2010A&A...524A..92C,2011ApJ...739..103A, Arimoto2020} and the external dissipation models \citep{Meszaros94, kumar09, Beloborodov14, fraija17}, although the latter may have difficulty in explaining the correlated temporal behavior of the GeV emission and keV--MeV emission in some GRBs~\citep{tang17}. Even in the internal dissipation models, it is not clear yet from which mechanism the extra PL component arises from. As shown in previous literature, either the photopion production or the Bethe-Heitler pair production of relativistic protons can reproduce the additional spectrum component at GeV band. Beside the hadronic origin model, the inverse Compton scattering of high-energy electrons can also reproduce such a spectral feature~\citep{wang18}. Therefore, it may be difficult to reveal the origin of the extra component solely from the GeV observations. As will be also discussed in this study, observations at lower energies may provide a clue to differentiate these models.

The rest of the paper is organized as follows.
In \S 2, we perform the spectral analysis of selected GRBs. In \S 3, the spectral fitting results are presented and discussed. In \S 4, we discuss the possible origin of the PL spectral component. The conclusions are presented in \S 5.

\section{Methodology} \label{sec:Data Analysis}

\subsection{Sample Selection}
A main criterion employed to select our sample is that the high-energy photons need to be detected by the \textit{Fermi}$-$LAT instrument during the GBM $T_{90}$ interval, among which 90\% of the burst fluence (50-300~keV) was accumulated. With a contemporaneous detection of the LAT and the GBM, we thus can perform the broadband spectral analysis between GBM $T_{05}$ and GBM $T_{95}$, which are the start and the end of GBM $T_{90}$.

Among 17 short bursts presented in the LAT catalog, we exclude 5 GRBs with no high-energy photons detected above 100~MeV during GBM $T_{90}$ intervals, i.e., GRB 090531B, GRB 110529A, GRB 160829A, GRB 170127C and GRB 180703B. Moreover, we also exclude GRB 160702A, since its GBM data is not archived in the GBM catalog.
Furthermore, we include in our sample a short burst, GRB 190515A~\citep{2019GCN.24560....1K}, that satisfies our selection criterion and was detected after the LAT catalog time period, namely between August 2018 and August 2020.
Finally, we also include the long GRB 160709A, although both catalogs classify it as a long burst. Indeed, \cite{2019ApJ...876...76T} classify it as a short hard GRB. In the spectral analysis, we only consider the main bursting phase of GRB 160709A, ranging from 0.32 and $\sim$0.77 s post trigger time, as discussed in ~\cite{2019ApJ...876...76T}.

Our sample includes 13 SGRBs from August 2008 to August 2020, which are listed in Table \ref{table:grbsample}, where the GBM trigger time ($T_{0}$ in Mission Elapsed Time, MET), $T_{90}$, $T_{05}$, $T_{95}$ are reported. Positions reported by the LAT catalog are employed for the LAT data reduction, as shown in Table~\ref{table:grbsample}.

\begin{deluxetable*}{ccccccrrc}
\tablecaption{Durations and positions of 13 GRBs in our sample}
\label{table:grbsample}
\tablehead{
\colhead{GRB}
&\colhead{GBM $T_{0}$\tablenotemark{a}}
&\colhead{GBM $T_{90}$}
&\colhead{GBM $T_{05}$}
&\colhead{GBM $T_{95}$}
&\colhead{LLE Detection\tablenotemark{b}}
&\colhead{LAT RA.\tablenotemark{c}}
&\colhead{LAT Decl.\tablenotemark{c}}
&\colhead{LAT Ref.\tablenotemark{d}}\\
\colhead{}
&\colhead{s}
&\colhead{s}
&\colhead{s}
&\colhead{}
&\colhead{}
&\colhead{}
&\colhead{}
&\colhead{}
}
\startdata
081024B	&	246576161.864	&	0.640	&	-0.064	&	0.576	&	Yes	&	323.01	&	20.84	&	(1)	\\
081102B	&	247308301.506	&	1.728	&	-0.064	&	1.664	&	--	&	212.95	&	30.33	&	...	\\
090227B	&	257452263.407	&	0.304	&	-0.016	&	0.288	&	Yes	&	11.80	&	32.20	&	...	\\
090228A	&	257489602.911	&	0.448	&	0	&	0.448	&	--	&	98.60	&	-28.79	&	...	\\
090510	&	263607781.971	&	0.960	&	-0.048	&	0.912	&	Yes	&	333.57	&	-26.62	&	...	\\
110728A	&	333508824.816	&	0.704	&	-0.128	&	0.576	&	--	&	173.57	&	4.34	&	...	\\
120830A	&	368003226.533	&	0.896	&	0	&	0.896	&	--	&	88.59	&	-28.79	&	...	\\
120915A	&	369360044.638	&	0.576	&	-0.320	&	0.256	&	--	&	240.95	&	57.04	&	...	\\
140402A	&	418090209.998	&	0.320	&	-0.128	&	0.192	&	--	&	207.66	&	5.97	&	...	\\
141113A	&	437559466.503	&	0.448	&	-0.064	&	0.384	&	--	&	182.32	&	77.38	&	...	\\
171011C	&	529442792.946	&	0.480	&	-0.448	&	0.032	&	--	&	168.48	&	10.03	&	...	\\
160709A	&	489786547.512	&	0.448	&	0.320	&	0.768	&	Yes	&	236.11	&	-28.51	&	(2)	\\
190515A	&	579587588.135	&	1.264	&	-0.112	&	1.152	&	--	&	137.69	&	29.28	&	(3)	\\
\enddata
\tablenotetext{a}{GBM burst trigger time in the format of the \textit{Fermi} Mission Elapsed Time}
\tablenotetext{b}{Yes indicates that \textit{Fermi}$-$LAT Low-Energy (LLE) data is available}
\tablenotetext{c}{Central position employed for the \textit{Fermi}$-$LAT detection}
\tablenotetext{d}{(1)\cite{2019ApJ...878...52A}; (2)\cite{2019ApJ...876...76T}; (3)\cite{2019GCN.24560....1K}}
\tablenotetext{e}{For GRB 160709A, selected time range is the main prompt GRB emission phase reported in \cite{2019ApJ...876...76T}}
\end{deluxetable*}

\subsection{Event Selection and Background Estimation}
\textit{Fermi}$-$GBM and \textit{Fermi}$-$LAT data are used in our spectral analysis. For 4 GRBs as shown in Table \ref{table:grbsample}, \textit{Fermi}$-$LAT Low-Energy (LLE) data are also combined in the spectral fitting. All data are available in the High Energy Astrophysics Science Archive Research Center~\footnote{https://fermi.gsfc.nasa.gov/ssc/data/access/}.

\textbf{GBM data.} For each GRB, we select three NaI detectors most close to the GRB position and one BGO detector with the lowest angle of incidence, which are presented in Table \ref{table:detector}. We analyze NaI time-tagged event (TTE) data with energy between 8 and 900 keV as well as BGO TTE data with energy between 250 keV and 40 MeV, excluding the overflow channels. The GBM backgrounds are usually estimated by fitting the observed TTE data tens of seconds before and after the source emission intervals. Because of the short durations ($<$ 2 seconds) in our sample, it is found that two time intervals are reasonable to derive a good count-rate background for the selected GRB detectors by the auto-determined polynomial order fitting, such as [-25,-10] and [15,30] away from the GBM trigger time. Instrument response files are selected with the \textit{rsp2} files, however, if no \textit{rsp2} files are included in the archived GBM data for some GRBs, such as GRB 120830A, GRB 120915A, GRB 140402A and GRB 141113A, the \textit{rsp} files are selected since our spectral analysis is performed for the GRBs with the relative short durations\citep{2014ApJS..211...13V,2016ApJS..223...28N}.

\begin{deluxetable*}{c|cc|cc|cc|cc}
\tablecaption{Information for the selected GBM detectors}
\label{table:detector}
\tablehead{
\colhead{GRB name}
&\colhead{1$^{st}$ NaI}
&\colhead{D\tablenotemark{a}}
&\colhead{2$^{nd}$ NaI}
&\colhead{D\tablenotemark{a}}
&\colhead{3$^{rd}$ NaI}
&\colhead{D\tablenotemark{a}}
&\colhead{BGO}
&\colhead{D\tablenotemark{a}}\\
\colhead{}
&\colhead{}
&\colhead{degree}
&\colhead{}
&\colhead{degree}
&\colhead{}
&\colhead{degree}
&\colhead{}
&\colhead{degree}
}
\startdata
081024B	&	n0	&	30.98	&	n6	&	23.56	&	n9	&	27.67	&	b1	&	73.66	\\
081102B	&	n0	&	30.66	&	n1	&	18.99	&	n5	&	44.20	&	b0	&	45.46	\\
090227B	&	n1	&	27.27	&	n2	&	19.31	&	n5	&	51.05	&	b0	&	54.22	\\
090228A	&	n0	&	10.35	&	n1	&	24.73	&	n3	&	40.40	&	b0	&	67.17	\\
090510	&	n0	&	34.17	&	n6	&	7.07	&	n7	&	32.68	&	b1	&	81.59	\\
110728A	&	n0	&	31.01	&	n1	&	33.34	&	n9	&	28.14	&	b1	&	87.21	\\
120830A	&	n0	&	22.98	&	n1	&	21.41	&	n3	&	39.17	&	b0	&	53.13	\\
120915A	&	n0	&	17.40	&	n3	&	36.17	&	n6	&	28.08	&	b0	&	79.00	\\
140402A	&	n0	&	28.28	&	n3	&	32.60	&	n6	&	20.26	&	b0	&	84.18	\\
141113A	&	n3	&	33.64	&	n6	&	38.46	&	n7	&	33.70	&	b1	&	89.78	\\
160709A	&	n3	&	13.06	&	n4	&	42.93	&	n6	&	44.79	&	b0	&	70.29	\\
171011C	&	n0	&	25.21	&	n1	&	23.77	&	n3	&	37.08	&	b0	&	51.29	\\
190515A	&	n0	&	38.36	&	n1	&	43.30	&	n9	&	18.98	&	b1	&	77.38	\\
\enddata
\tablenotetext{a}{Angular separation between the pointing of the GBM detector and the GRB position in unit of degree}
\end{deluxetable*}

\textbf{LLE data.} There are 4 GRBs in our sample with the LLE detection as shown in Table \ref{table:grbsample}, such as GRB 081024B, GRB 090227B, GRB 090510 and GRB 160709A. Events with energy between 20~MeV and 100~MeV are selected in our spectral analysis. Reduction of the LLE data is same as that of the GBM data when estimating the background.

\textbf{LAT data.} LAT--\textit{Transient020E} events with a zenith angle cut of 100$^o$ are selected for each burst, whose energy are between 100~MeV to 10~GeV. For GRB 090510, the highest photon energy is about 30~GeV, thus the maximum energy is 100~GeV. Region of interest (ROI) is chosen within the radius of 12$^o$ from the localization report in Table~\ref{table:grbsample}.

\begin{figure*}
\centering
\includegraphics[width=0.70\textwidth]{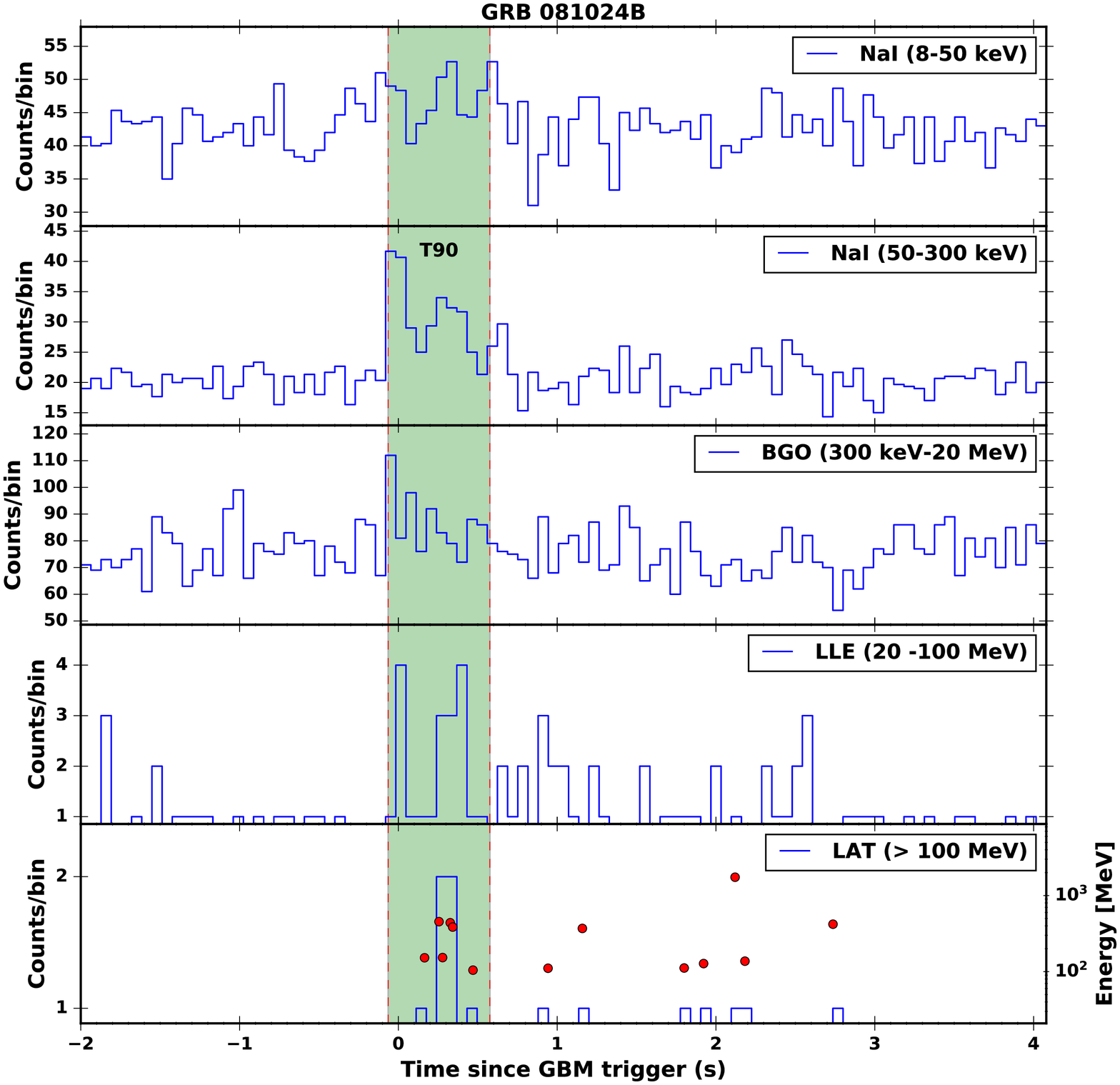}
\caption{Composite light curve for GRB 081024B. The average count-rate of three NaI detectors (the first two panels), BGO (the third panel), LLE (the fourth panel), and LAT Transient-class events above 100 MeV within a 12$^\circ$ ROI (bottom panel). The shadowed region is the selected time intervals to be analyzed.The filled circles are events that have a probability $>$0.9 of being associated with the GRB.}
\label{fig:lc1}
\end{figure*}

After the event selection, the count-rate light curve is built for each GRB. For example, the composite light curve for GRB 081024B is shown in Figure ~\ref{fig:lc1}.

\subsection{Fitting models}

In order to test the existence of the additional PL spectral component, 6 typical empirical functions as the fitting models are employed to fit the broadband gamma-ray data of each GRB, which are described below:

(i) The blackbody function (BB), which is usually modified by the Planck spectrum and given by the photon flux
\begin{equation}
\frac{dN}{dE}=A_{\rm BB} \frac{E^{2}}{\exp[E/kT]-1},
\label{Eq:BB}
\end{equation}
where $k$ is the Boltzmann's constant, and the joint parameter $kT$ as a output parameter in common. It is found the peak energy in $E^2 dN/dE$ spectrum of the BB is about 3.92 times of the value of $kT$, that is $E_{p, {\rm BB}}\approx 3.92 kT$.
In all the functions here and below, A is the normalization constant.

(ii) The Band function (BAND), which is written same as that in \citep{1993ApJ...413..281B},
\begin{eqnarray}
\frac{dN}{dE} = A_{\rm BAND} \left\{ \begin{array}{ll}
(\frac{E}{100\ {\rm keV}})^{\alpha} e^{[-E(2+\alpha)/E_{p}]}, & E\leq \frac{\alpha-\beta}{2+\alpha}E_{p} \\
\\
(\frac{(\alpha-\beta)E_{p}}{(2+\alpha) 100\ {\rm keV}})^{(\alpha-\beta)} e^{(\beta-\alpha)}(\frac{E}{100\ {\rm keV}})^{\beta}, & E\geq \frac{\alpha-\beta}{2+\alpha}E_{p} \\
\end{array} \right.
\label{Eq:BAND}
\end{eqnarray}
where $\alpha$, $\beta$ are the low-energy photon index and the high-energy photon index respectively, and $E_{p}$ is the peak energy in the $E^2 dN/dE$ spectrum, which is reported in the section of results as $E_{p, {\rm BAND}}$.

(iii) The cutoff power-law model (CPL), written as
\begin{equation}
\frac{dN}{dE} = A_{\rm CPL} (\frac{E}{100\ {\rm keV}})^{\alpha} e^{-E/E_{\rm c}},
\label{Eq:CPL}
\end{equation}
where $\alpha$ is the photon index and $E_{\rm c}$ is the cutoff energy, the peak energy in $E^2 dN/dE$ spectrum for the CPL ($E_{p, {\rm CPL}}$) equals to $(2+\alpha)E_{c}$, say, $E_{p, {\rm CPL}} = (2+\alpha)E_{c}$.

(iv) The composition function of the BB and a simple power law function (BB+PL), that is
\begin{equation}
\frac{dN}{dE} = (\frac{dN}{dE})_{\rm BB} + A_{\rm PL} (\frac{E}{100\ {\rm keV}})^{\Gamma_{\rm PL}},
\label{Eq:BBPL}
\end{equation}
where $(\frac{dN}{dE})_{\rm BB}$ is same as Eq. \ref{Eq:BB} and the $\Gamma_{\rm PL}$ is the photon index of the PL function.

(v) The composition function of the BAND and a simple power law function (BAND+PL), that is
\begin{equation}
\frac{dN}{dE} = (\frac{dN}{dE})_{\rm BAND} + A_{\rm PL}(\frac{E}{100\ {\rm keV}})^{\Gamma_{\rm PL}},
\label{Eq:BANDPL}
\end{equation}
where $(\frac{dN}{dE})_{\rm BAND}$ is same as Eq. \ref{Eq:BAND} and the $\Gamma_{\rm PL}$ is the photon index of the PL function.

(vi) The composition function of the CPL and a simple power law function (CPL+PL), that is
\begin{equation}
\frac{dN}{dE} = (\frac{dN}{dE})_{\rm CPL} + A_{\rm PL}(\frac{E}{100\ {\rm keV}})^{\Gamma_{\rm PL}},
\label{Eq:CPLPL}
\end{equation}
where $(\frac{dN}{dE})_{\rm CPL}$ is same as Eq. \ref{Eq:CPL} and the $\Gamma_{\rm PL}$ is the photon index of the PL function.

\subsection{Spectral fitting and the best-fitting model selection}

In this work, we perform the Markov chain Monte Carlo (MCMC) fitting technique based on the bayesian statistic by using the the Multi-Mission Maximum Likelihood package (3ML; \citealt{2015arXiv150708343V}), to carry out all the spectral analysis and the parameter estimation, which requires the corresponding informative priors and the posterior sampling of parameter space in each fitting model.

\subsubsection{Informative priors selection}

The informative priors are adopted by using the typical spectral parameters from the Fermi-GBM catalog~\citep{2021ApJ...913...60P}, hereafter we named it as the typical priors (TP). For all parameters in the TP scenario, we set the initial parameter values and the parameter range same as the default value in the 3ML package except for the normalization ($A$), whose lower-bound and upper-bound are calculated by $10^{-5}$ and $10^{5}$ times its initial value. Distributions of normalization ($A$) are the logarithm uniform distribution (LogU), the photon indices ($\alpha, \beta$ and $\Gamma$) are sharing the gaussian distributions (G) and the parameters in the unit of keV ($E_p, E_c$ and $kT$) are distributed in the logarithm normal distributions (LogN). For all gaussian distributions (G), the central value ($\mu$) equals to initial parameter value and the one standard deviation ($\sigma$) is fixed at 0.5. For all logarithm normal distributions (logN), both $\mu$ and $\sigma$ are at the initial parameter values. The TP scenario is accepted in several public work on spectral analysis of the \textit{Fermi}-GBM GRBs~\citep{2019ApJS..245....7L,2021ApJS..254...35L,2019ApJ...886...20Y}.
The details of these priors are presented in Table \ref{table:prior}. For the compositive models (BB+PL, BAND+PL and CPL+PL), we use the joint informative priors above. We also test the uniform priors (UP) for all spectral parameters, whose initial values and parameter ranges are same as that in the TP scenario but with the uniform parameter distributions. Results in the UP scenario are presented in the Appendix \ref{appendixUP}, which draw the conclusion that the resultant parameters in both scenarios are consistent with each other, therefore the results in the TP scenario are presented in the following sections.

\begin{deluxetable*}{cllll}
\tablecaption{Prior setting}
\label{table:prior}
\tablehead{
\colhead{Function}
&\colhead{Parameter}
&\colhead{Initial Value}
&\colhead{Parameter Range}
&\colhead{TP scenario\tablenotemark{a}}
}
\startdata
PL	&	$A$, $\Gamma$	&	$10^{-4}$, -2.0	&	[$10^{-9},10$], [-10.0,10.0]	&	logU, G	\\
BB	&	$A$, $kT$	&	$10^{-4}$, 30	&	[$10^{-9},10$], [0,$10^{5}$]	&	logU, LogN	\\
BAND	&	$A$, $\alpha$, $\beta$, $E_p$	&	$10^{-4}$, -1.0, -2.0, 500	&	[$10^{-9},10$], [-1.5,3.0], [-5.0,-1.6], [0,$10^{7}$]	 &	 logU, G, G, LogN\\
CPL	&	$A$, $\alpha$, $E_c$	&	$10^{-4}$, -2.0, 30	&	[$10^{-9},10$], [-10.0,10.0], [0,$10^{7}$]	&	logU, G, LogN\\
\enddata
\tablenotetext{a}{For the typical priors (TP), LogU represents the logarithm uniform distribution, G represents the gaussian distribution, LogN is the logarithm normal distribution.}
\end{deluxetable*}

\subsubsection{Posterior sampling and the best-fitting model selection}\label{bestmodelselection}

We employ the $emcee$, a sampling method included in the 3ML package, to sample the posterior, which is an extensive, pure-python implementation of Goodman \& Weare's Affine Invariant MCMC Ensemble sampler~\citep{goodman10}. The $emcee$ use multiple walkers to explore the parameter space of the posterior. For each sampling, we set the number of chains (walkers) to be 20, the number of learning samples to be 3000 that we do not include in the final results, and the number of global samples to be 15000. Twice MCMC fittings are performed, one with the initial parameter values, and the other one with the resultant median parameter values.

In order to know which of a suite of models best represents the data, two information criterion are usually presented to choose the best-fitting model for our sampling SGRBs, such as the Akaike Information Criterion (AIC; \citealt{1974ITAC...19..716A}) and Bayesian Information Criterion (BIC; \citealt{1978AnSta...6..461S}). Here we prefer the BIC to select a best-fitting model due to the large sampling in our MCMC fittings. Given any two estimated models, the preferred model is the one that provides the smaller BIC value. Here we use the difference in BIC value ($\Delta$BIC = BIC$_{\rm model\ B}$ $-$ BIC$_{\rm model\ A}$) to describe the evidence against a candidate model (model B) to the best model (model A) in the model comparisons. If $\Delta$BIC is larger than 10, the evidence against the candidate model is very strong \citep{kass1995bayes}.

\section{Results}

\subsection{Best-fitting models}
Comparison results of different models of 13 SGRBs are presented in Table \ref{table:comparison}. We identify 3 subclasses according to the best fit models: 8 GRBs are best fitted by the BB+PL model (Class A), 4 GRBs by the CPL+PL model (Class B) and GRB 090510 by the BAND+PL model (Class C).
The Spectral energy distributions (SEDs) for 3 GRBs from each subclass, namely GRB 081024B, GRB 090227B and GRB 090510, are plotted together with the marginal posterior distributions in Figure \ref{fig:bestfit}, while SEDs for other 10 GRBs are shown in Appendix \ref{fig:bestfitleft1}. In all SEDs, we calculated the residual values by $(f_d-f_m)^2/\sigma_{f_d}^2$, where $f_d$, $\sigma_{f_d}$ are the binned observational \textit{Fermi} data and the corresponding 1$\sigma$ errors, $f_m$ are the fluxes calculated by the best-fitting models. All residuals in 13 GRBs are between $0$ and $3.0$, which imply the good spectral fittings for all GRBs. All resultant parameters of the best-fitting models are presented in Table \ref{table:bestfit}.

Hereafter, we categorize the BB, BAND and CPL functions as the main component and the PL function as the extra PL component. The extra PL component is present in all 13 analyzed SGRBs, which might imply the common existence of an extra energy dissipation process in SGRBs.

\begin{deluxetable*}{ccccccc|c}
\tablecaption{$\Delta{\rm BIC}$ and the best-fitting models}
\label{table:comparison}
\tablehead{
\colhead{GRB}
&\colhead{BB}
&\colhead{BAND}
&\colhead{CPL}
&\colhead{BB+PL}
&\colhead{BAND+PL}
&\colhead{CPL+PL}
&\colhead{Best Model\tablenotemark{a}}
}
\startdata
081024B	&	$>$10\tablenotemark{b}	&	$>$10	&	$>$10	&	0	&	$>$10	&	$>$10	&	BB+PL	\\
081102B	&	$>$10	&	$>$10	&	$>$10	&	0	&	$>$10	&	$>$10	&	BB+PL	\\
090227B	&	$>$10	&	$>$10	&	$>$10	&	$>$10	&	$>$10	&	0	&	CPL+PL	\\
090228A	&	$>$10	&	$>$10	&	$>$10	&	$>$10	&	3	&	0	&	CPL+PL	\\
090510	&	$>$10	&	$>$10	&	$>$10	&	$>$10	&	0	&	$>$10	&	BAND+PL	\\
110728A	&	$>$10	&	$>$10	&	$>$10	&	0	&	$>$10	&	$>$10	&	BB+PL	\\
120830A	&	$>$10	&	$>$10	&	$>$10	&	$>$10	&	$>$10	&	0	&	CPL+PL	\\
120915A	&	$>$10	&	$>$10	&	$>$10	&	0	&	$>$10	&	$>$10	&	BB+PL	\\
140402A	&	$>$10	&	$>$10	&	$>$10	&	0	&	$>$10	&	$>$10	&	BB+PL	\\
141113A	&	$>$10	&	$>$10	&	$>$10	&	0	&	$>$10	&	$>$10	&	BB+PL	\\
160709A	&	$>$10	&	$>$10	&	$>$10	&	$>$10	&	3	&	0	&	CPL+PL	\\
171011C	&	$>$10	&	$>$10	&	$>$10	&	0	&	$>$10	&	$>$10	&	BB+PL	\\
190515A	&	$>$10	&	$>$10	&	$>$10	&	0	&	$>$10	&	$>$10	&	BB+PL	\\
\enddata
\tablenotetext{a}{Best-fitting model with the $\Delta{\rm BIC} = 0$}
\tablenotetext{b}{$>$10 represents the best model against this candidate model.}
\end{deluxetable*}

\begin{figure}[p]
	\centering
	\subfigure[]{
		\includegraphics[width=0.45\textwidth]{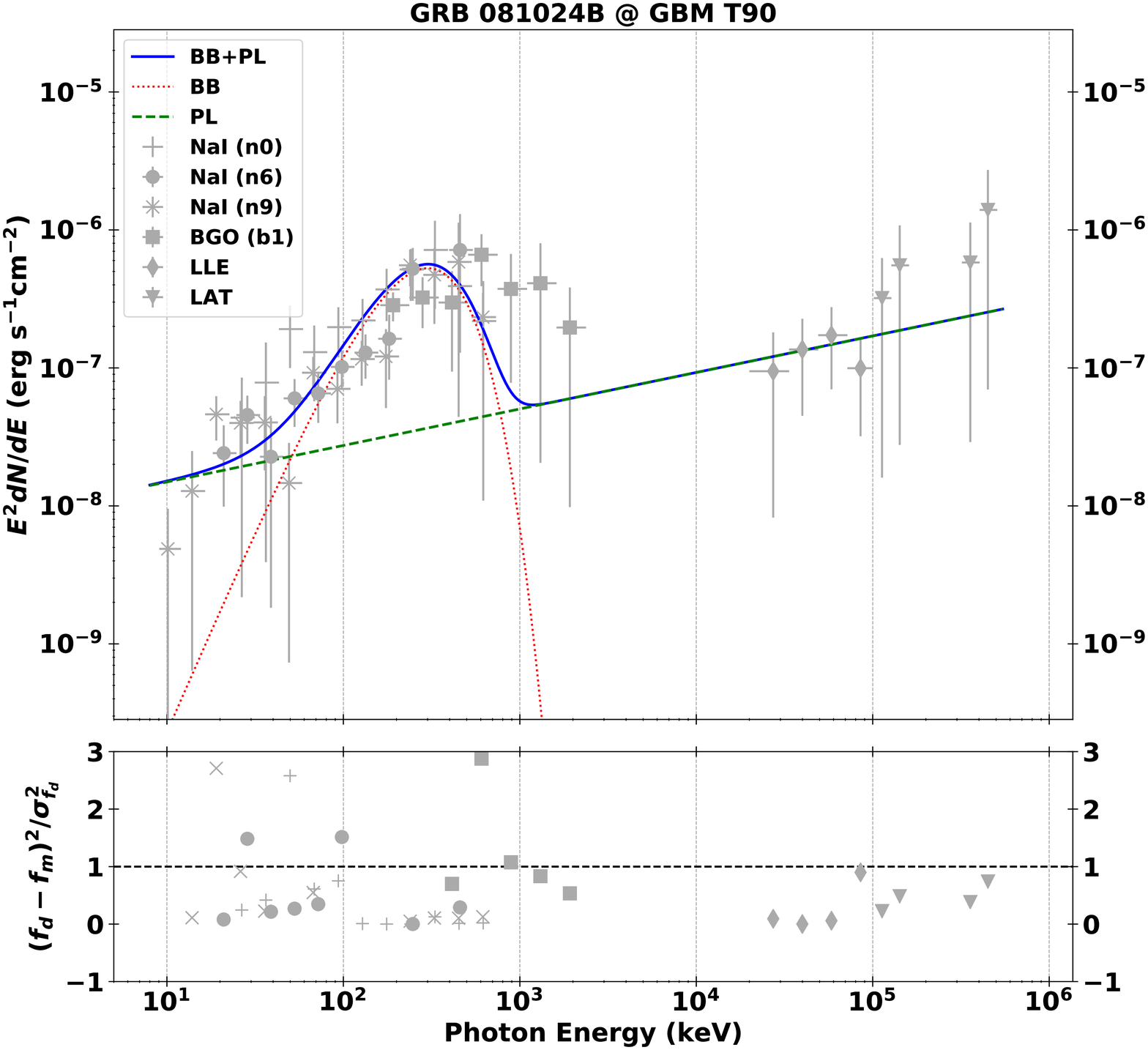}}
	\subfigure[]{
		\includegraphics[width=0.35\textwidth]{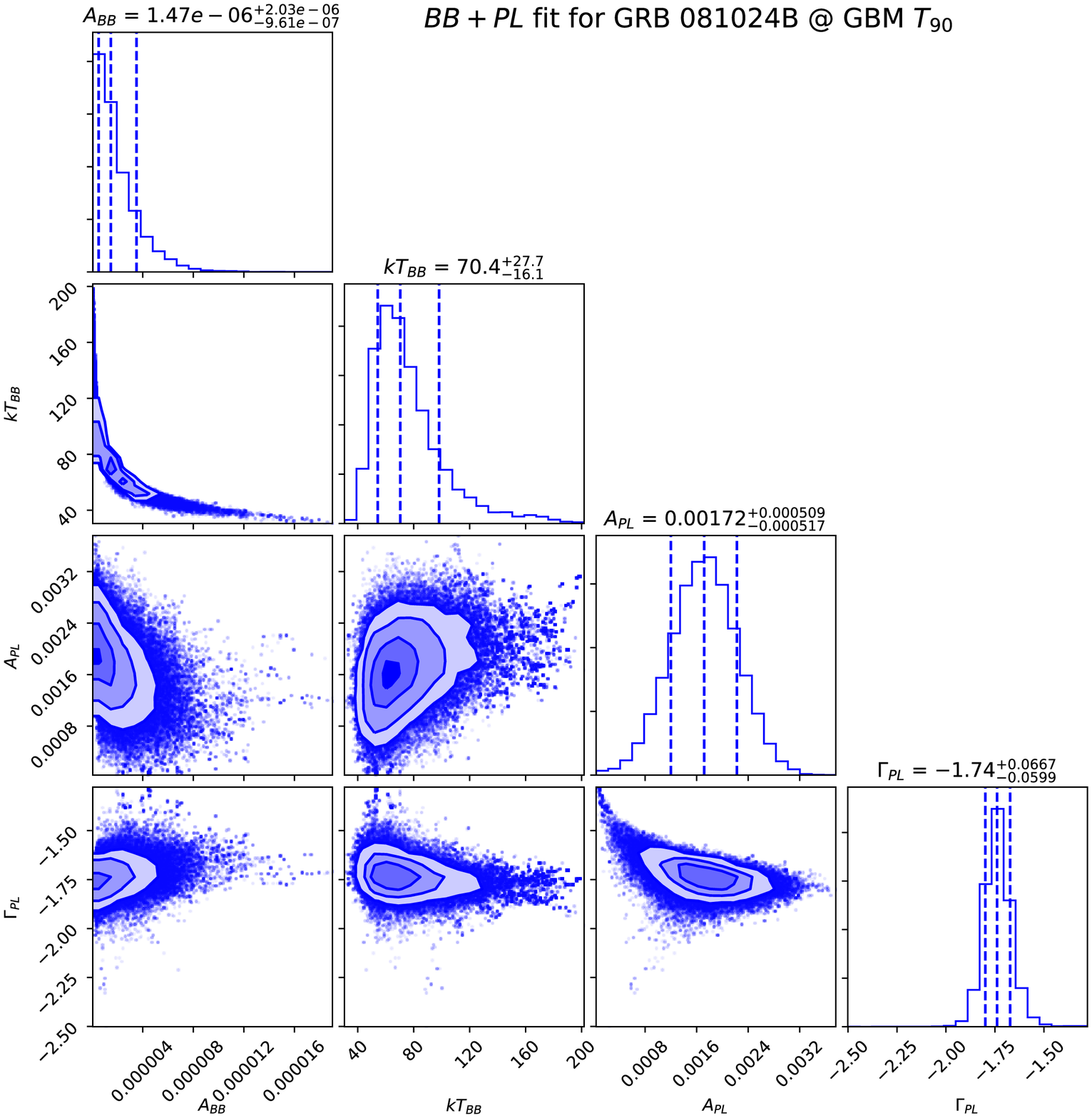}}
	\subfigure[]{
		\includegraphics[width=0.45\textwidth]{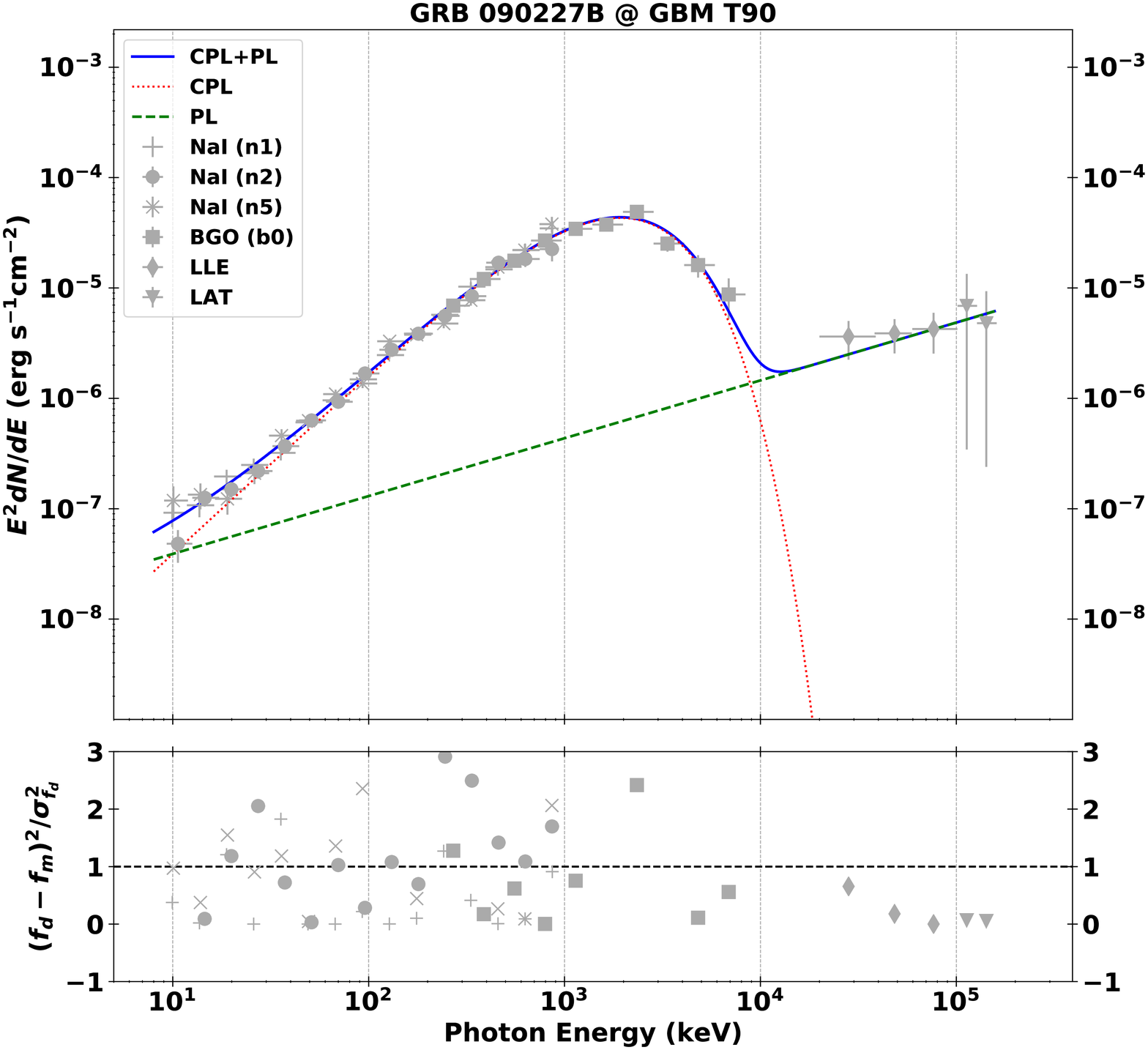}}
	\subfigure[]{
		\includegraphics[width=0.35\textwidth]{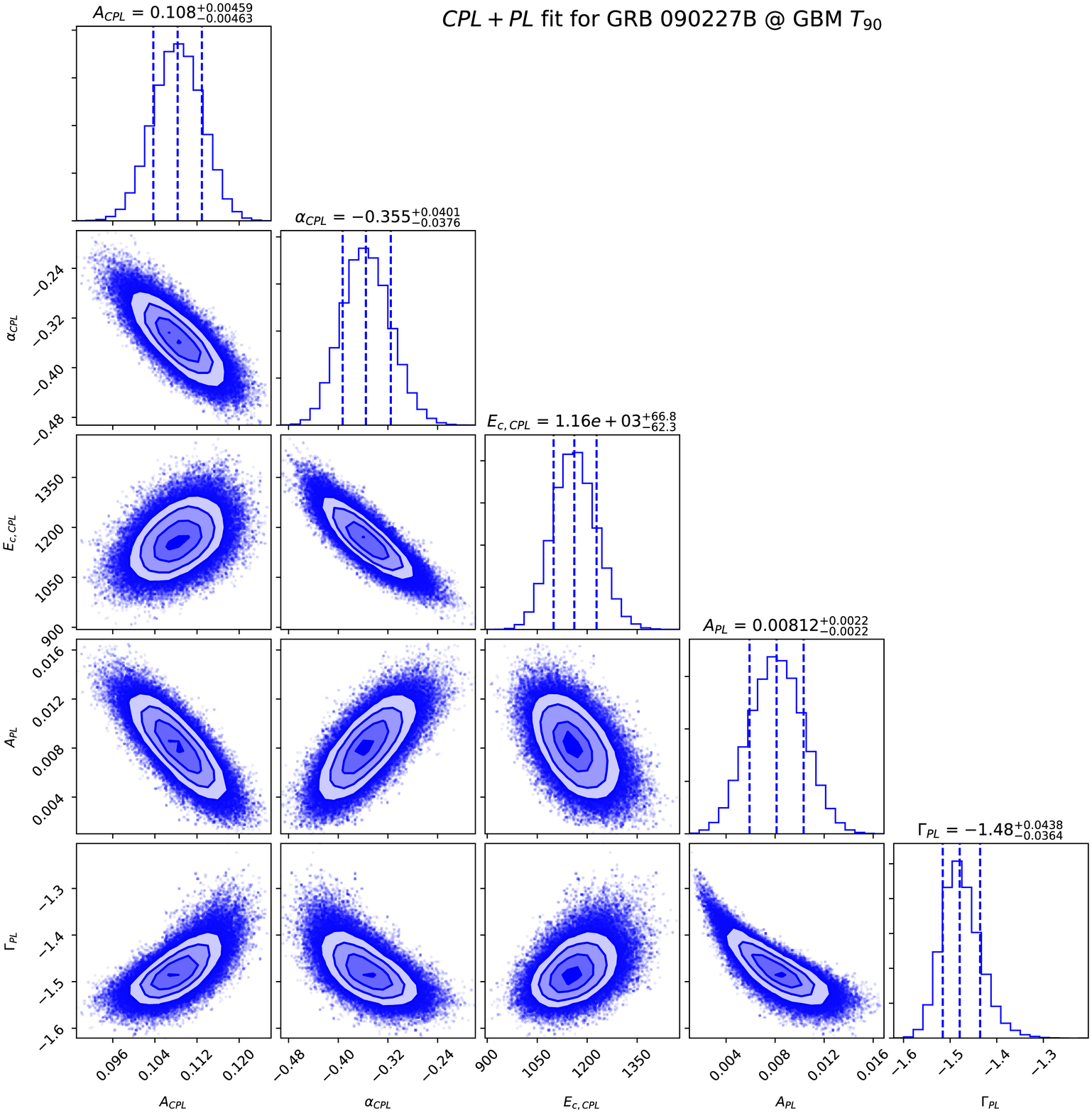}}
	\subfigure[]{
		\includegraphics[width=0.45\textwidth]{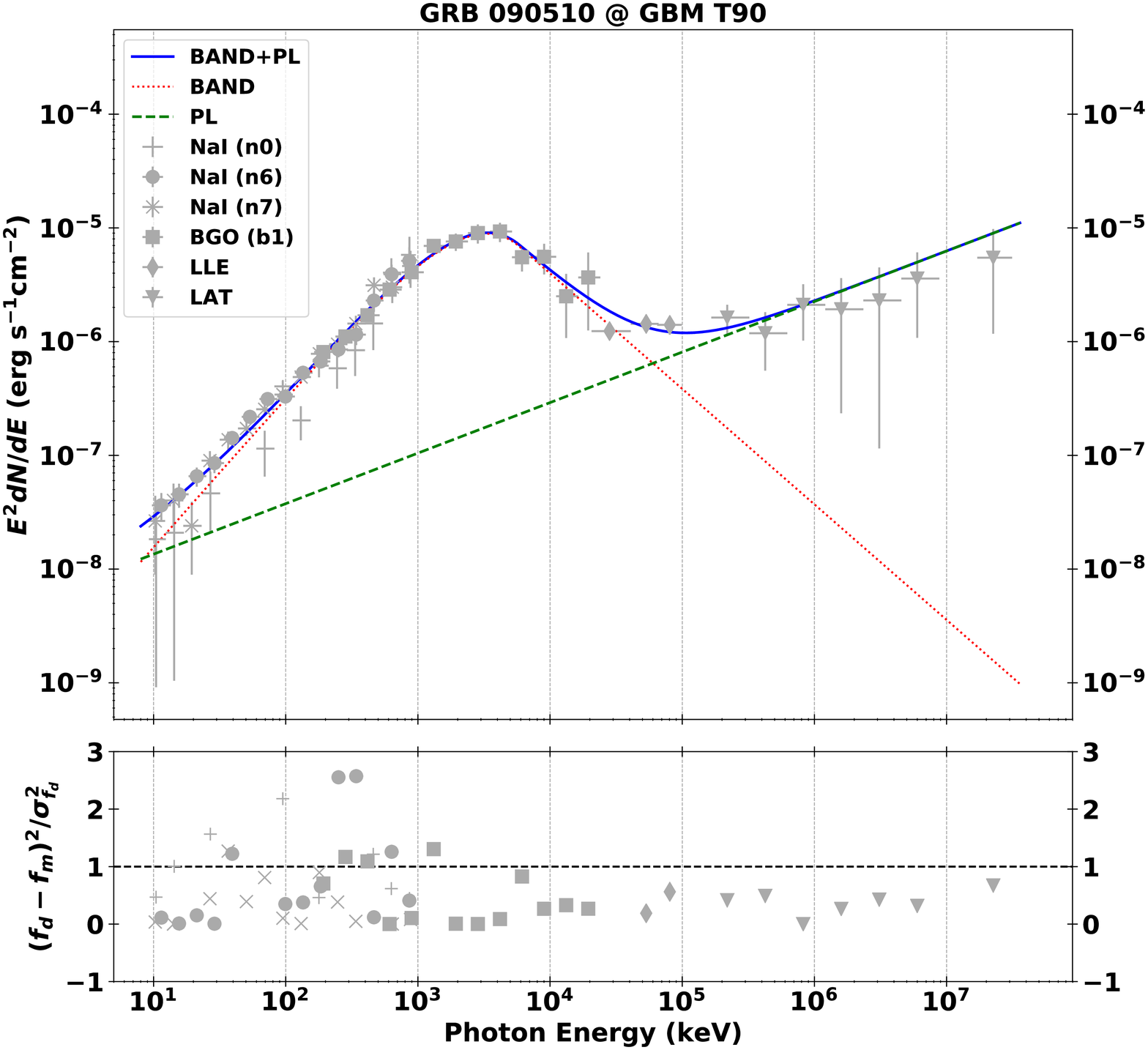}}
	\subfigure[]{
		\includegraphics[width=0.35\textwidth]{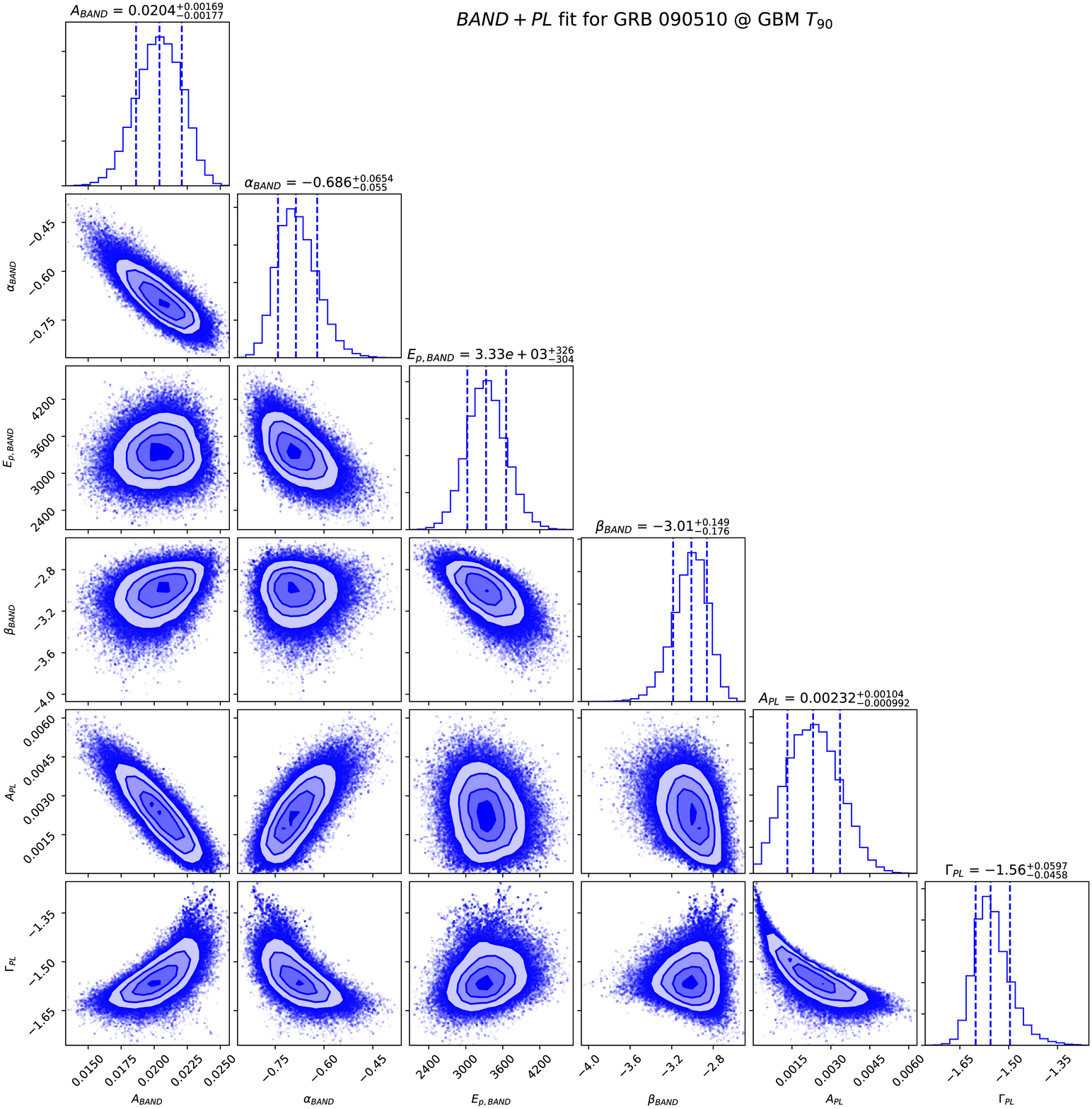}}
	\caption{Best-fitting spectral energy distributions (SEDs) and the marginal posterior distributions for SGRBs in our sample. In each SED (a, c, e), gray points are the binned observational data by \textit{Fermi}, the red dotted line is the modeled main component, the green dashed line is the modeled extra PL component and the blue solid line represents the sum of both components.}
\label{fig:bestfit}
\end{figure}

\begin{deluxetable*}{c|cccc|ccc}
\label{table:bestfit}
\tablecaption{Derived parameters of the best-fitting model in the TP scenario}
\tablehead{
\colhead{Class}
&\colhead{Main component}
&\colhead{}
&\colhead{}
&\colhead{}
&\colhead{Extra PL component}
&\colhead{}
&\colhead{}
}
\startdata
BB+PL	&	$A_{\rm BB}$\tablenotemark{a}	&	  &	$E_{\rm p,BB}$\tablenotemark{b}	&	$F_{\rm BB}$\tablenotemark{c}	&	 $A_{\rm PL}$\tablenotemark{d}	&	$\Gamma_{\rm PL}$	&	$F_{\rm PL}$\tablenotemark{e}	\\
081024B	&	19.9	$\pm$	15.0	&				&	301	$\pm$	86	&	7.2	$\pm$	5.4	&	171.0	$\pm$	51.3	&	 -1.74	$\pm$	0.06	&	9.1	$\pm$	2.7	\\
081102B	&	13.1	$\pm$	4.5	&				&	297	$\pm$	32	&	4.4	$\pm$	1.5	&	161.0	$\pm$	41.8	&	 -1.85	 $\pm$	0.12	&	5.4	$\pm$	1.4	\\
110728A	&	23.6	$\pm$	11.5	&				&	244	$\pm$	40	&	3.7	$\pm$	1.8	&	17.9	$\pm$	17.7	&	 -1.93	$\pm$	0.42	&	0.4	$\pm$	0.4	\\
120915A	&	12.5	$\pm$	4.5	&				&	347	$\pm$	42	&	8.0	$\pm$	2.9	&	9.9	$\pm$	9.6	&	-1.89	 $\pm$	 0.42	&	0.2	$\pm$	0.2	\\
140402A	&	3.8	$\pm$	1.4	&				&	588	$\pm$	79	&	20.0	$\pm$	7.5	&	19.8	$\pm$	18.7	&	 -2.13	 $\pm$	0.52	&	0.3	$\pm$	0.2	\\
141113A	&	3.0	$\pm$	1.6	&				&	539	$\pm$	111	&	11.3	$\pm$	5.9	&	58.7	$\pm$	53.7	&	 -1.87	 $\pm$	0.35	&	1.7	$\pm$	1.6	\\
171011C	&	23.8	$\pm$	3.5	&				&	209	$\pm$	46	&	69.7	$\pm$	1.5	&	17.2	$\pm$	14.8	 &	 -1.90	$\pm$	0.46	&	0.4	$\pm$	0.4	\\
190515A	&	2.4	$\pm$	0.6	&				&	679	$\pm$	114	&	22.2	$\pm$	5.9	&	49.8	$\pm$	40.3	&	 -1.83	 $\pm$	0.25	&	1.7	$\pm$	1.4	\\
\hline
CPL+PL	&	$A_{\rm CPL}$\tablenotemark{a}	&	 $\alpha_{\rm CPL}$ &	$E_{\rm p,CPL}$\tablenotemark{b}	&	$F_{\rm CPL}$\tablenotemark{c}	&	 $A_{\rm PL}$\tablenotemark{d}	&	$\Gamma_{\rm PL}$	&	$F_{\rm PL}$\tablenotemark{e}	 \\
090227B	&	10.8	$\pm$	0.5	&	-0.35	$\pm$	0.04	&	1915	$\pm$	106	&	886.0	$\pm$	37.7	&	813.0	 $\pm$	220.0	&	-1.48	$\pm$	0.04	&	106.0	$\pm$	28.7	\\
090228A	&	10.0	$\pm$	0.4	&	-0.27	$\pm$	0.09	&	767	$\pm$	85	&	192.0	$\pm$	8.7	&	438.0	$\pm$	 238.0	&	-2.06	$\pm$	0.22	&	6.0	$\pm$	3.3	\\
120830A	&	2.0	$\pm$	0.1	&	-0.16	$\pm$	0.11	&	1005	$\pm$	159	&	67.2	$\pm$	4.1	&	23.8	$\pm$	 23.5	&	-2.02	$\pm$	0.43	&	0.4	$\pm$	0.4	\\
160709A	&	2.6	$\pm$	0.2	&	-0.13	$\pm$	0.08	&	1784	$\pm$	180	&	269.0	$\pm$	23.1	&	380.0	 $\pm$	 100.0	&	-1.66	$\pm$	0.05	&	24.7	$\pm$	6.5	\\
\hline
BAND+PL	&	$A_{\rm BAND}$\tablenotemark{a}	&	 $\alpha_{\rm BAND}$  &	$E_{\rm p,BAND}$\tablenotemark{b}	&	$F_{\rm BAND}$\tablenotemark{c}	 &	$A_{\rm PL}$\tablenotemark{d}	&	$\Gamma_{\rm PL}$	&	$F_{\rm PL}$\tablenotemark{e}	 \\
&		&	($\beta_{\rm BAND}$)	&		&		&		&		&		\\
090510	&	2.0	$\pm$	0.2	&	-0.68	$\pm$	0.06	&	3322	$\pm$	316	&	241.0	$\pm$	20.9	&	235.0	 $\pm$	 103.0	&	-1.56	$\pm$	0.05	&	229.0	$\pm$	101.0	\\
&		&	(-3.02	$\pm$	0.16)	&		&		&		&		&		\\
\enddata
\tablenotetext{a}{Normalizations for the main components, $A_{\rm BB}$ in unit of $10^{-7} {\rm ph\ keV^{-1} cm^{-2}\ s^{-1}}$, $A_{\rm CPL}$ and $A_{\rm BAND}$ in unit of $10^{-2} {\rm ph\ keV^{-1} cm^{-2}\ s^{-1}}$}
\tablenotetext{b}{Peak energy of the $E^2 dN/dE$ spectrum in unit of {\rm keV}}
\tablenotetext{c}{Fluxes of the main components in unit of $10^{-7} {\rm erg\ cm^{-2}\ s^{-1}}$}
\tablenotetext{d}{Normalization for the extra component in unit of $10^{-5} {\rm ph\ keV^{-1} cm^{-2}\ s^{-1}}$}
\tablenotetext{e}{Flux of the extra PL component in unit of $10^{-7} {\rm erg\ cm^{-2}\ s^{-1}}$}
\end{deluxetable*}

\subsection{Parameter Distributions}
\begin{figure}
\centering
\includegraphics[width=0.60\textwidth]{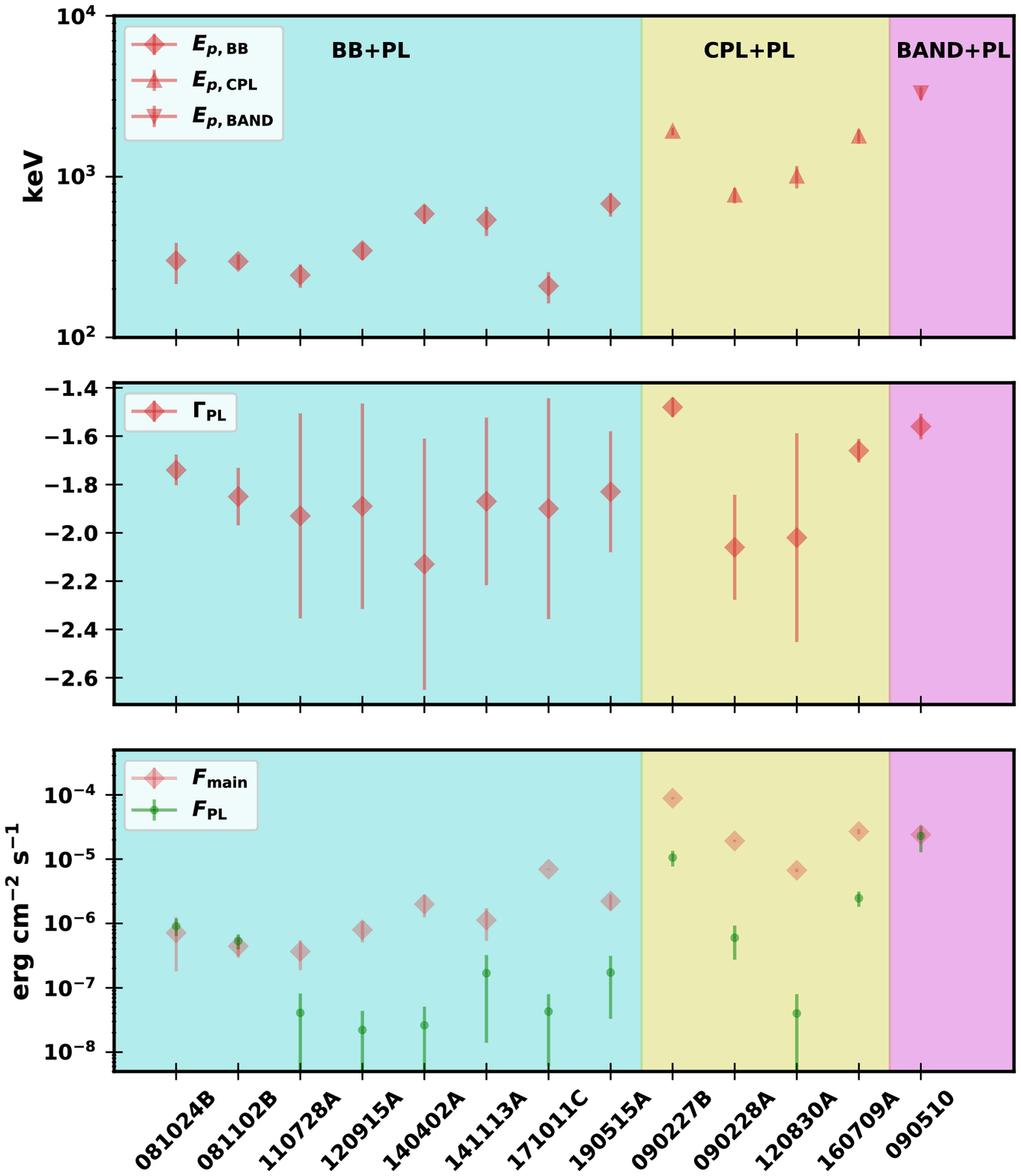}
\caption{Top: distributions of $E_{\rm p, BB}$, $E_{\rm p, CPL}$ and $E_{\rm p, BAND}$. Middle : distribution of indices of the extra PL component ($\Gamma_{\rm PL}$). Bottom: distributions of the average flux of both components ($F_{\rm main}$ and $F_{\rm PL}$). The cyan shadows are for the class of BB+PL, the yellow shadows for the class of the CPL+PL and the magenta shadows for the class of the BAND+PL.}
\label{fig:distributions}
\end{figure}

For the main components, we calculate the peak energy ($E_{\rm p}$) in the $E^2 dN/dE$ spectrum. For the standard BB component, the peak energy is found about 3.92 times of $kT_{\rm BB}$ in our sample, that is $E_{\rm p, BB} \approx 3.92\ kT_{\rm BB}$, which is also employed in \citet{2020ApJ...902...40Z} and \citet{2019ApJ...876...76T}. For the CPL component, the peak energy is calculated as $E_{\rm p, CPL} = (2+\alpha)E_c$. The values of the peak energy ($E_{\rm p, BB}$, $E_{\rm p, CPL}$ and $E_{\rm p, BAND}$) are reported in Table \ref{table:bestfit}.
As shown in Figure \ref{fig:distributions}, $E_{\rm p}$ is ranging from $\sim$ 200 keV to $\sim$ 3 MeV. It is found that the peak energies of the main components of GRBs that are best-fitted with CPL or BAND are larger than those best-fitted with BB.

For the extra components, the observed spectra are generally hard, with the spectral index ($\Gamma_{\rm PL}$) ranging from  $\sim$ -2.1 to -1.5, e.g., 10 out of 13 GRBs in our sample with central values of $\Gamma_{\rm PL}$ larger than -2.0. Note that it does not necessarily mean the absence of a softer PL component in reality, because GRBs with softer PL components may not be detectable to \textit{Fermi}$-$LAT.

\subsection{Correlation between $F_{\rm main}$ and $F_{\rm PL}$}
Spectral fluxes between ${\rm 8\ keV}$ and $E_{\rm max}$ (the maximum photon energy detected by \textit{Fermi}$-$LAT) are calculated by integrating the $E dN/dE$ spectrum, denoted by  $F_{\rm main}$ for the main component and $F_{\rm PL}$ for the extra PL component. Then we test the correlation between them by the linear fit in the logarithm space, such as
\begin{equation}
\log F_{\rm PL} = m + n\log F_{\rm main}
\end{equation}
where $m$ and $n$ are the free parameters. This fitting is performed by the basic linear regression analysis in the popular $Origin$ scientific package, which can give the coefficient of determination ($R^2$, $0 < R^2 <1$). For the linear fit, two variables, such as $F_{\rm main}$ and $F_{\rm PL}$ in our work, are positively correlated if the Pearson-correlation coefficient ($R$, $-1 < R <1$) is close to $1$.

\begin{figure}
\centering
\includegraphics[width=0.45\textwidth]{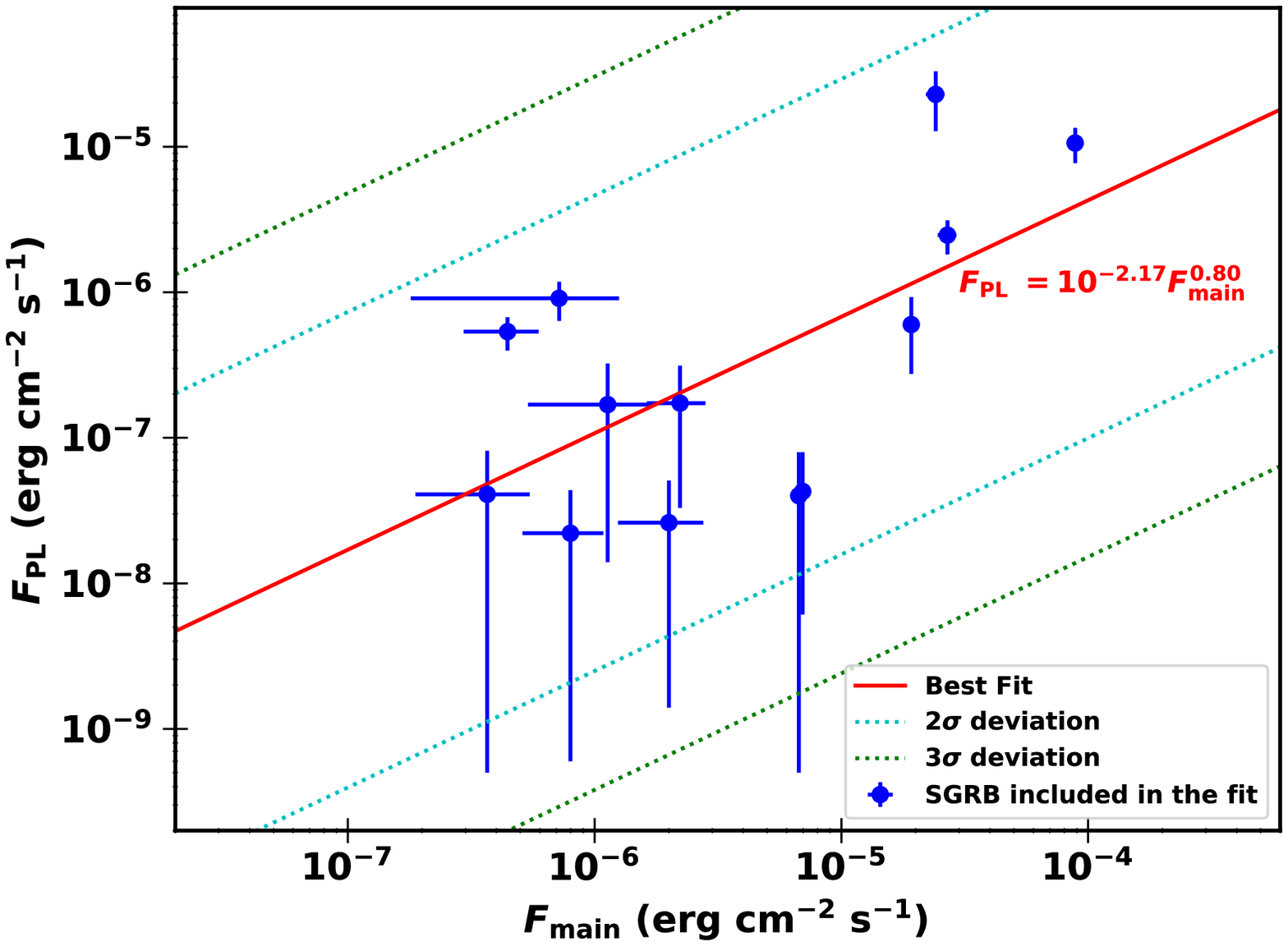}
\includegraphics[width=0.45\textwidth]{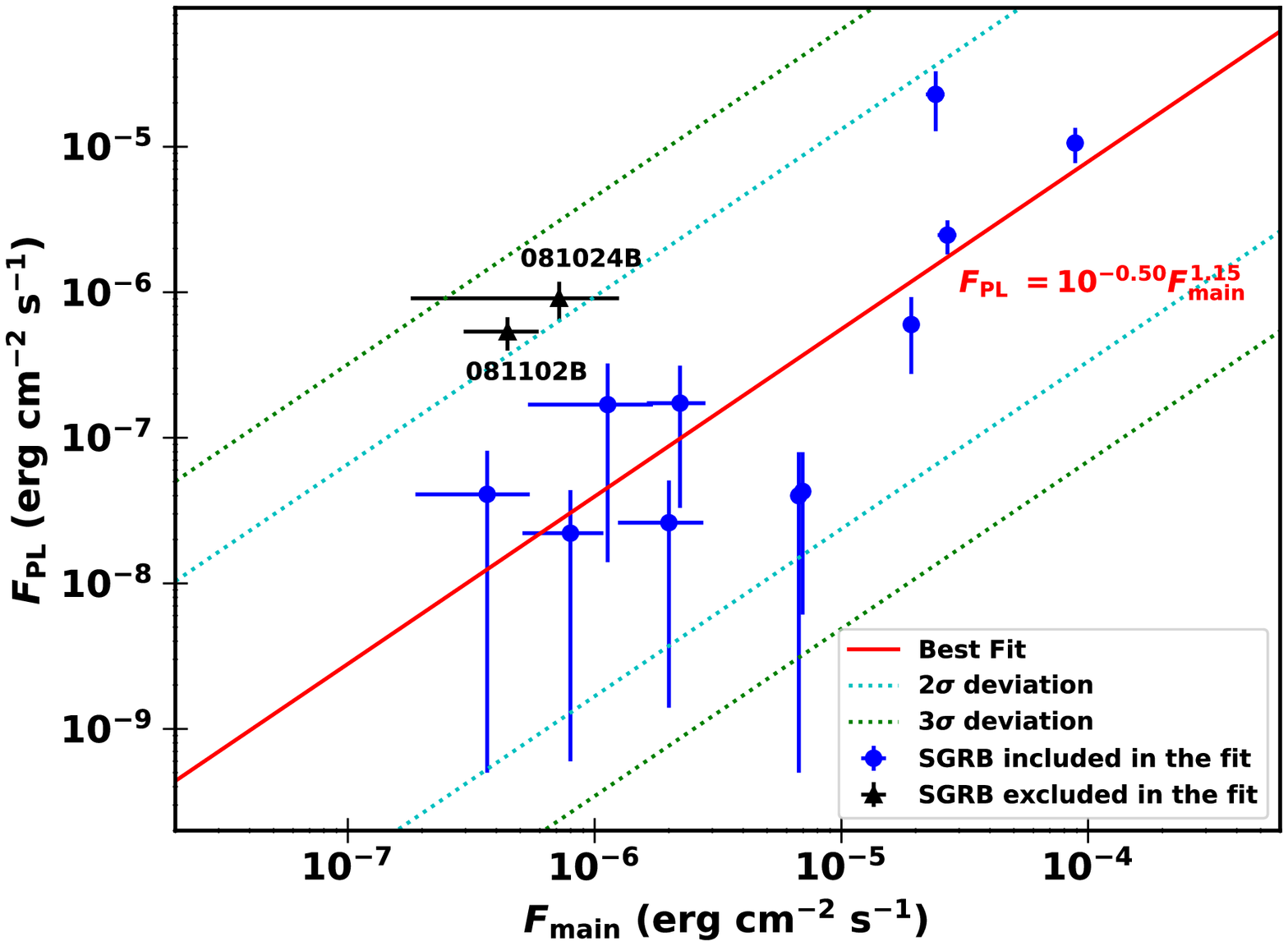}
\caption{Left: Linear fit for $F_{\rm main}$ and $F_{\rm PL}$ in the logarithm space for all GRBs in our sample. Right: Linear fit for $F_{\rm main}$ and $F_{\rm PL}$ in the logarithm space excluding GRB 081024B and GRB 081102B. The solid line is the best fit, the cyan and green dotted lines represent the 2 $\sigma$, 3$\sigma$ deviations from the best fit respectively.}
\label{fig:correlation}
\end{figure}

We found a moderate correlation between $F_{\rm main}$ and $F_{\rm PL}$ for all GRBs in our sample, with $R$ = 0.62, $m$ = $-2.17\pm1.67$ and $n$ = $0.80\pm0.31$. The best fit for the correlation is written as
\begin{equation}
\log F_{\rm PL} = 10^{-2.17\pm1.67} + (0.80\pm0.31) \log F_{\rm main},
\label{eq:correlation1}
\end{equation}
where both $F_{\rm PL}$ and $F_{\rm main}$ are in unit of ${\rm erg\ cm^{-2} s^{-1}}$. This correlation is plotted in the left panel of Figure \ref{fig:correlation}, in which two GRB, GRB 081024B and 081102B, are far from the best-fitting line comparing with other GRBs.
Therefore, a similar linear fit is performed by excluding GRB 081024B and 081102B. In this case, $F_{\rm main}$ and $F_{\rm PL}$ has a stronger positive correlation than that in Equation \ref{eq:correlation1}, with $R$ = 0.80,  $m$ = $-0.50\pm1.55$ and $n$ = $1.15\pm0.29$, which is presented as
\begin{equation}
\log F_{\rm PL} = 10^{-0.50\pm1.55} + (1.15\pm0.29) \log F_{\rm main},
\label{eq:correlation2}
\end{equation}
which is shown in the right panel of Figure \ref{fig:correlation}.
In this strong positive correlation, GRB 081024B and 081102B deviate from the correlation at about $3\sigma$ level, with an excessively high ratio of the PL component to the main component with respect to the ratios in other GRBs.
This requires an efficient conversion of the jet's kinetic energy to the nonthermal particles in the prompt emission phase of the GRB or implies an important contribution from the early afterglow. In the latter case, it requires an early deceleration of the GRB jet by the interstellar medium, probably caused by a high initial bulk Lorentz factor $\Gamma_0$ for the jet given the deceleration timescale being $t_{\rm dec}=0.5 (n_{\rm ISM}/1\rm cm^{-3})^{1/3}(E_k/10^{50}\rm erg)^{1/3}(\Gamma_0/1000)^{-8/3}\,$s~\citep{Meszaros94}.

\section{Possible origin of the extra PL components}

In order to explore the possible origins of the extra PL components, we need to understand two main features of this spectral component, namely, the spectral slope which is found to approximately range between $[-2.0, -1.5]$ (see the middle panel of Figure~\ref{fig:distributions}) and the flux amplitude relative to that of the main spectral component (see the bottom panel of Figure~\ref{fig:distributions}). The origin of extra high-energy emission (especially above 100 MeV) is still under debate. The late-time and long-lasting high-energy gamma-ray emission from GRBs, such as 080916C, 090510, and 090902B, may arise from afterglow emission rather than the prompt emission \citep{kumar09, ghisellini10, razzaque11}. However, the high-energy emission in the early stage presents a rapid variability and a temporal correlation with the keV/MeV emission, implying an internal dissipation origin \citep{maxham11, tang17}. The extra high-energy emission, usually detected by Fermi-LAT in the brightest GRBs, has been discussed about its origin via various high-energy processes, such as Comptonized thermal, self-schchrotron Compton (SSC) \citep{rees94, asano07}, proton-induced cascade \citep{vietri97, dermer06, 2009ApJ...705L.191A, wang18} and proton synchrotron emission \citep{totani98}. Except the one-zone model, multi-zone leptonic models including the SSC scenario\citep{corsi10, daigne11}, the external inverse Compoton scenario\citep{toma11, peer12}, and the synchrotron radiation scenario \citep{ioka10} have been invoked as well.

Generally, the extra PL component (including high-energy part) implies the presence of nonthermal relativistic particles accelerated during the prompt emission phase, which can arise when a multiplicative stochastic process (that reaches lognormal in equilibrium) is truncated before equilibrium is achieved \citep{Mitzenmacher04,Reed04,Fang12}, such as the Fermi-type acceleration processes. Here we consider two possible one-zone scenarios: one is the product of hadronic interactions of accelerated protons (or the hadronic model), and the other is the inverse Compton (IC) radiation of accelerated electrons (or the leptonic model). In either scenario, the produced high-energy photons will be likely absorbed by the radiation from the main component as the GRB fireball is quite compact. The secondary electron/positron pairs will be generated and radiate new generation of photons via the IC process in the radiation field and via the synchrotron process in the magnetic field. The new generation of photons will repeat the above process unless the energies of the new-generated photons drops below the threshold of the pair production process. Such a process is called the
electromagnetic (EM) cascade. It will largely modify the spectrum of the initially generated high-energy gamma rays and dominate the PL component. To deal with the EM cascade process, we follow the treatment described by \cite{wang18}. Note that we do not aim to explain the main spectral component, so we simply treat it as a target photon field for $\gamma\gamma$ annihilation and the IC radiation. For the main spectral component, although most of GRB prompt emission spectra around keV-MeV are present as a non-thermal shape and usually can be modeled as a smoothly broken power law, i.e., the BAND function, a thermal emission originating from the photosphere is a natural prediction of the generic fireball scenario \citep{paczynski86,shemi90,meszaros93,peer12,hascoet13}. The relative strength of thermal emission and non-thermal emission should depends on the various environments \citep{daigne02,ryde05}. For the SGRBs samples in this paper, the main spectral components of most of them can be described better by a BB emission. The detailed origin of the main spectral component is beyond the scope of this paper, here we only approximate it to be a BB emission in the calculation although in some GRBs the main spectral components are found to be best described with CPL or BAND.

We consider a GRB located at $z=1$ with a bulk Lorentz factor of $\Gamma=300$ and a dissipation radius $R=10^{14}\,\rm cm$. The main spectral component is assumed to be a diluted BB distribution with a temperature of $kT=100(1+z)\,\mathrm{keV}$ and an isotropic-equivalent luminosity of $L_{\rm BB}=6\times10^{52}\rm erg/s$.

\begin{figure}
\centering
\includegraphics[width=0.45\textwidth]{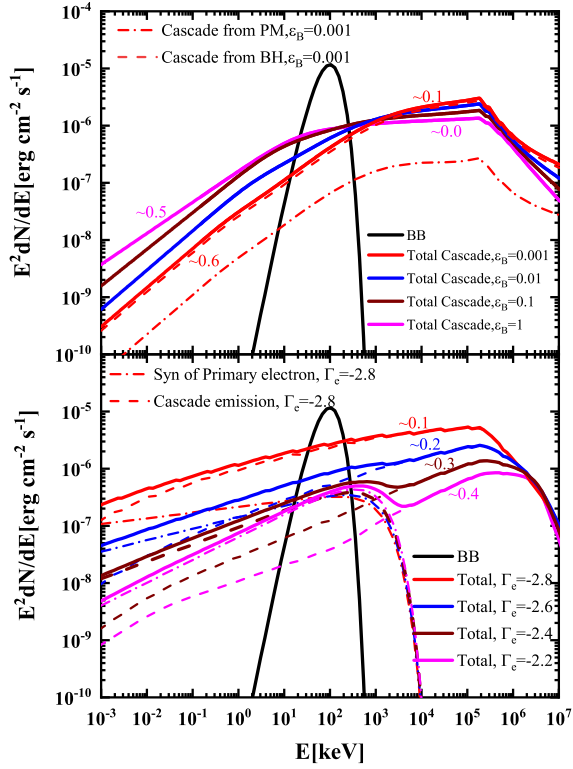}
\caption{The extra components for the hadronic model (top) and leptonic model (bottom). In the bottom panel, the solid lines is the sum of the cascade emission (dashed lines) and the synchrotron radiation (dash-dotted lines) of primary electrons. The approximate slopes for different lines are present with same colors.}
\label{fig:theoretical}
\end{figure}

For the hadronic model, the radiation at GeV energies is \textbf{dominanted} by the electromagnetic cascade initiated by the hadronic processes, including the photomeson (PM) process and the Bethe-Heitler (BH) process. Protons can be accelerated at dissipation radius by some processes, e.g., internal shocks or magnetic reconnections. In this case, we define a magnetic equipartition coefficient ($\varepsilon_B$) as the ratio of the magnetic field energy density $U_B$ to the photon energy density of the BB component $U_{bb}$, i.e., ${\varepsilon _B}=U_B/U_{\rm bb}$. The proton spectrum is assumed to be a power-law distribution with a slope of $p=-2$ and a maximum proton energy $E_{p,\rm max}>0.15{\rm GeV}\Gamma^2/kT \approx 10^{17}\,\rm eV$ in order to have an efficient photomeson process. The isotropic-equivalent luminosity for protons is taken to be $6\times 10^{53}\,\rm erg/s$ corresponding to a baryon loading factor of 10. The accelerated protons can generate high-energy gamma-rays and electrons through PM and BH processes and then initiate the EM cascade in the photon field and the magnetic filed. As shown in \cite{wang18}, different values of ${\varepsilon_B}$ can lead to different indexes of cascade emission due to the different ratio between the contributions from the synchrotron radiation and that of the IC radiation. Indeed, as we can see in the top panel of Figure~\ref{fig:theoretical}, for a larger ${\varepsilon _B}$, the photon index is close to $-2.0$, while for a smaller ${\varepsilon _B}$, the photon index tends to be larger. The photon index of the cascade emission in the $1-10$\,keV energy range is about $-1.5$ in all the cases because it is mainly produced by the electrons cooled from higher energies and hence a $E^{-2}$ spectrum is expected for these cooled electrons~\citep{wang18}.

For the leptonic model, some electrons in the GRB fireball, in addition to those responsible for the main spectral component, are assumed be accelerated up to ultrarelativistic energies with a distribution of a power law $dN/dE = A{E^{  {\Gamma _e}}}$. The IC scattering on both the BB component and the synchrotron radiation of these ultrarelativistic electrons themselves can give rise to high-energy radiation. Similar to that in the hadronic model, the produced high-energy radiation will trigger an EM cascade. The relative contribution from the synchrotron process and the IC process of the cascade emission depends on the equipartition coefficient $\varepsilon_B$ in the same way shown in the hadronic scenario. So here we mainly explore the influence of the injection spectral index $\Gamma_e$ in the bottom panel of Figure~\ref{fig:theoretical} while fix $\varepsilon_B=0.01$, the flux of synchrotron radiations of primary electrons at $100\,\rm keV$ and the maximum electron Lorentz factor emitting a typical photon energy $\sim 1\,\rm MeV$. For $\Gamma_e=-2.8$, the spectral shape of the extra component is quite flat power-law with a photon index $\sim -1.9$, and for a larger $\Gamma_e$, the spectra become harder with a larger photon index.

\begin{figure}
\centering
\includegraphics[width=0.55\textwidth]{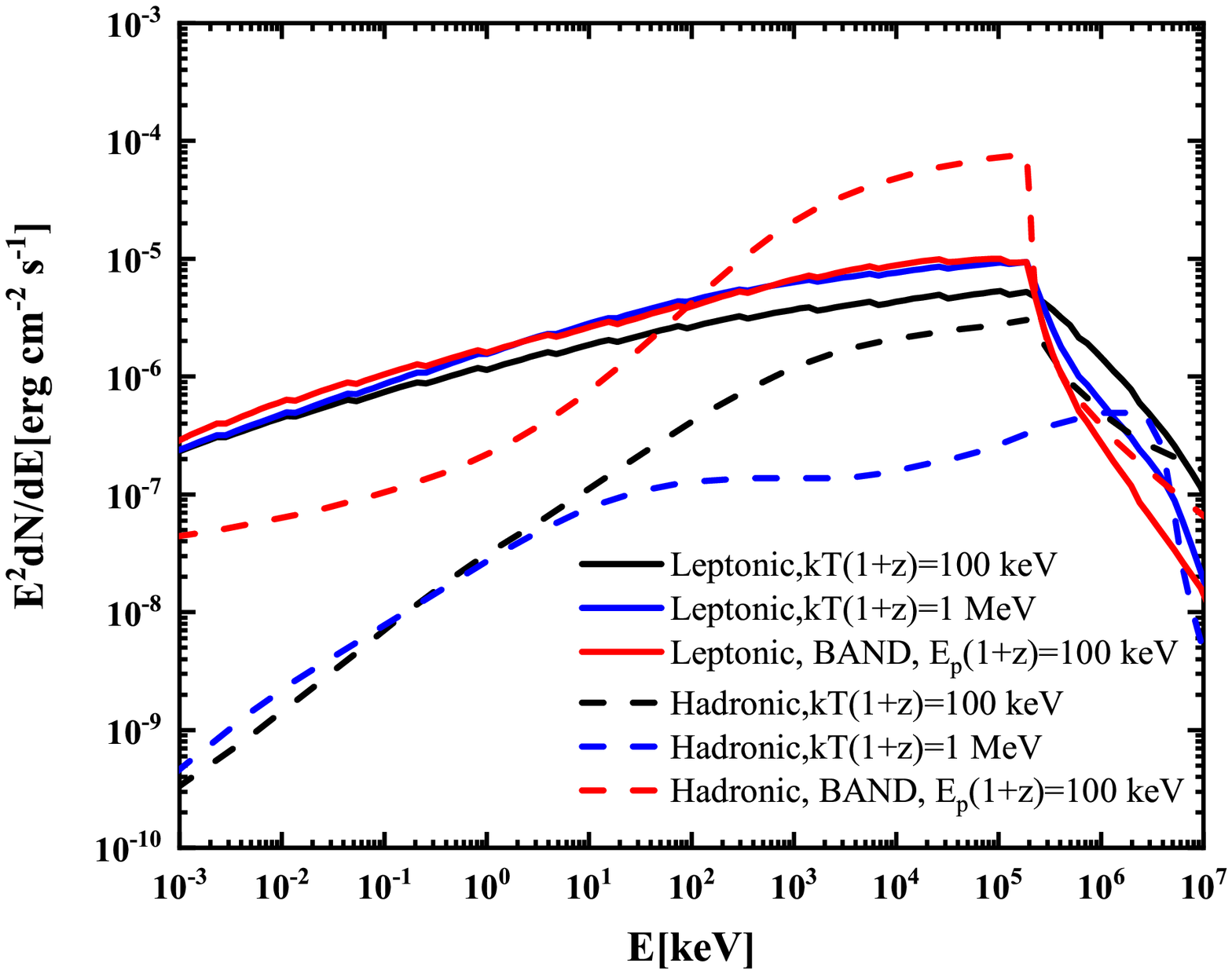}
\caption{The effects for different background photon distributions. The hadronic model (solid lines) and the leptonic model (dashed lines) are shown with two temperatures (black and blue) for the BB background photon distribution and with a BAND function (red) background photon distribution. For the different temperature of the BB component, the peak flux is involved with the same value of $10^{-5}\rm erg \,cm^{-2}\, s^{-1}$. The parameters of BAND component are $\alpha=-1.0$, $\alpha=-2.2$, and the same flux $10^{-5}\rm erg \,cm^{-2}\, s^{-1}$ as BB component. $\epsilon_B=0.01$ and $\Gamma_e=-2.8$ are adopted for all lines. The other parameters are same with the red lines in Fig~\ref{fig:theoretical} for the hadronic model and the leptonic model, respectively (see text for details).}
\label{fig:varied}
\end{figure}

We have also checked the dependence of our results on the assumed model parameters, e.g., the temperature of the BB that spans two orders of magnitude in Figure~\ref{fig:distributions} and a background photon field with the BAND function distribution. The BAND function distribution with the typical values of the low-energy photon index $\alpha=-1.0$, the high-energy photon index $\beta=-2.2$, the peak energy $E_p (1+z)=100\rm keV$ and the peak flux $10^{-5}\rm erg \,cm^{-2}\, s^{-1}$ is adopted to replace the BB distribution. For exploring the dependence of different temperatures and background photon field distribution, the same electron distribution with a power law $dN/dE = A{E^{  {\Gamma _e}}}$ for simplicity (the detailed origin of the BAND component is beyond the scope of the paper) and ${\Gamma _e}=-2.8$ are assumed for the leptonic model, the same proton distribution as in Figure~\ref{fig:theoretical} for the hadronic model and the same $\epsilon_B=0.01$ is adopted for both models. The final cascade emission depends on whether the EM cascade is fully developed and the total low-energy photon field including the initial photon field (BB or BAND distribution) and the cascade emission in keV-MeV energy range. In the GRB environment, the EM cascade is likely fully developed due to the relatively high photon density. As shown in Figure~\ref{fig:varied}, for the leptonic model, different temperatures (black and blue solid lines) would produce similar radiations because the EM cascade is fully developed and the low-energy photons from keV to MeV energies are approximately dominated by the high-flux cascade emission, and for the BAND function distribution as the background photon filed, the cascade emission shows a similar spectral shape and the magnitude of the cascade flux depends principally on the background photon filed and the adopted electron distribution. For the hadronic model in Figure~\ref{fig:varied}, since the low-energy photons ranging from keV to MeV energies from the cascade emission is weaker than that from the initial BB component, the initial BB component with the same peak flux and a higher temperature provides lower photon density, inducing a lower cascade emission.
When a BAND function distribution is involved in the hadronic model, a higher cascade flux is expected since we fixed the flux of BAND component as $10^{-5}\rm erg \,cm^{-2}\, s^{-1}$ same as that of the BB component and the low-energy photon index of BAND component, $-1.0$, is much smaller than that of BB component so that the BAND component provides much more photons with energies below $E_p$ than the BB component and make the hadronic processes (PM and BH) more efficient, inducing a higher injection luminosity for the cascade emission. At the ev-keV energy range, the radiations becomes flat (the red dashed line in Figure~\ref{fig:varied}) as the radiation below keV are mainly produced through the synchrotron radiation process by the electrons from the photon-photon annihilation rather than by the electrons cooled from higher energies (the latter one usually shows a typical fast cooling photon index, i.e., $\sim -1.5$). The BAND function distribution provides much more target photons with energies below $E_p$ and increases the photon-photon annihilation opacity. Except flattening at ev-keV energy range for the BAND function in the hadronic model, other characteristics of the spectral shapes for either the hadronic model or leptonic model in the different temperatures of the BB component or even treating the background photon field as the BAND function does not change significantly compared with those in Figure~\ref{fig:theoretical}. In addition, Even though taking such flattening at ev-keV energy range into account, the spectra for a quiet large energy range extending from eV to GeV, could be treated as a PL component approximately.

In summary, as shown from Figure~\ref{fig:theoretical}, both models can produce an approximate PL component ranging from keV to GeV energies within a certain range of index, which is consistent with our result of extra PL component for the SGRBs. However, for a flat PL component with a photon index close to $-2.0$, the low-energy excess up to 10 keV could be helpful to tell us which model is preferred since for the former one the photon index of low-energy excess is close to $-1.5$ while for the latter one it is $\sim (\Gamma_e-1)/2$. Nevertheless, the poor statistics at a few keVs makes it difficult to differentiate the two models with current observations. On the other hand, the cascade emissions of both models can extend down to the optical band, as shown in Figure~\ref{fig:theoretical}, and the flux difference at the optical band between the two models becomes distinct. Therefore, in the future, observations in the optical band of the prompt emission of GRBs may tell us which model is preferable. In addition, the hadronic model usually needs to invoke a relatively larger kinetic luminosity than leptonic model due to the lower radiation efficiency of protons than electrons and maybe exceed the typical energy budgets of GRBs. The hadronic model also naturally predicts the neutrino production which might be constrained by the stacking observation of IceCube as it was done in the case of LGRBs \citep{aartsen15,aartsen16,aartsen17}.

\section{Summary and Conclusion}

In this paper, we looked into the extra PL spectral components in short GRBs. By analyzing the combined \textit{Fermi}$-$GBM and LAT data, we identified the PL component in all 13 short GRBs in our sample, including GRB~090510 and GRB~160709A, whose extra PL component was already previously reported in literature. The average flux of the PL components within $T_{90}$ scale positively correlates with that of the main spectral components.  The slopes of the extra PL components of short GRBs distribute in the range between -2.0 to -1.5, which may be well reproduced by considering the electromagnetic cascade induced by ultra-relativistic protons or electrons accelerated in the prompt emission phase. In the future, observations with more statistics around keV energy band and the observations on the prompt optical GRB emission may tell us which model is preferable. In addition, the next-generation neutrino telescopes might play a key role in determining a preferred one of these two models.

\acknowledgments
We thank the anonymous referee who helped improve the manuscript, and the statistics editor for helpful suggestions. We are grateful to Yu Wang and Donggeun Tak for helpful discussions. This research made use of the High Energy Astrophysics Science Archive Research Center (HEASARC) Online Service at the NASA/Goddard Space Flight Center (GSFC). This work is supported by the NSFC under grants 11903017, 12065017, 12003007, U2031105 and 11975116, and the Fundamental Research Funds for the Central Universities (No. 2020kfyXJJS039).

\clearpage
\vspace{5mm}
\facilities{\textit{Fermi}$-$GBM, \textit{Fermi}$-$LAT}
\software{\textit{Fermi Science Tools}, \textit{Origin}, 3ML}

\clearpage

\appendix
\restartappendixnumbering

\section{Comparisons of results between the uniform priors (UP) scenario and the typical priors (TP) scenario}
\label{appendixUP}

In this section, we test the uniform-distribution priors for all parameters of 6 models, hereafter named as the UP. In the UP scenario, all normalizations ($A$), photon indices ($\alpha$, $\beta$ and $\Gamma$) and parameters of the break energy ($kT$,$E_c$ and $E_p$) are distributed in the uniform, which employed the same range as that in the typical priors (TP) scenario.

The best-fitting models can be shown in Table \ref{table:comparisonUP}. Of 13 GRBs with $\Delta{\rm BIC}$ equals 0 (Best Model), 10 GRBs are preferring with one-component model and 3 GRBs are with two-component model. There are several candidate models with $\Delta{\rm BIC}$ less then 10, which cannot be rejected by best-model-selection method described in Section \ref{bestmodelselection}. Therefore, in the UP scenario, we divided the best-fitting models into the best-fitting one-component models (Best 1C Model) and the best-fitting two-component models (Best 2C Model).

In order to compare results in the UP scenario with that in the TP scenario, we thus selected the Best 2C model. As shown in Table \ref{table:comparisonUP}, the Best 2C model of each GRB in the UP scenario is same as the best model of the corresponding GRB in the TP scenario. The resultant parameters in the Best 2C model are reported in Table \ref{table:bestfitUP}.

\begin{figure}[p]
	\centering
	\subfigure[]{
		\includegraphics[width=0.45\textwidth]{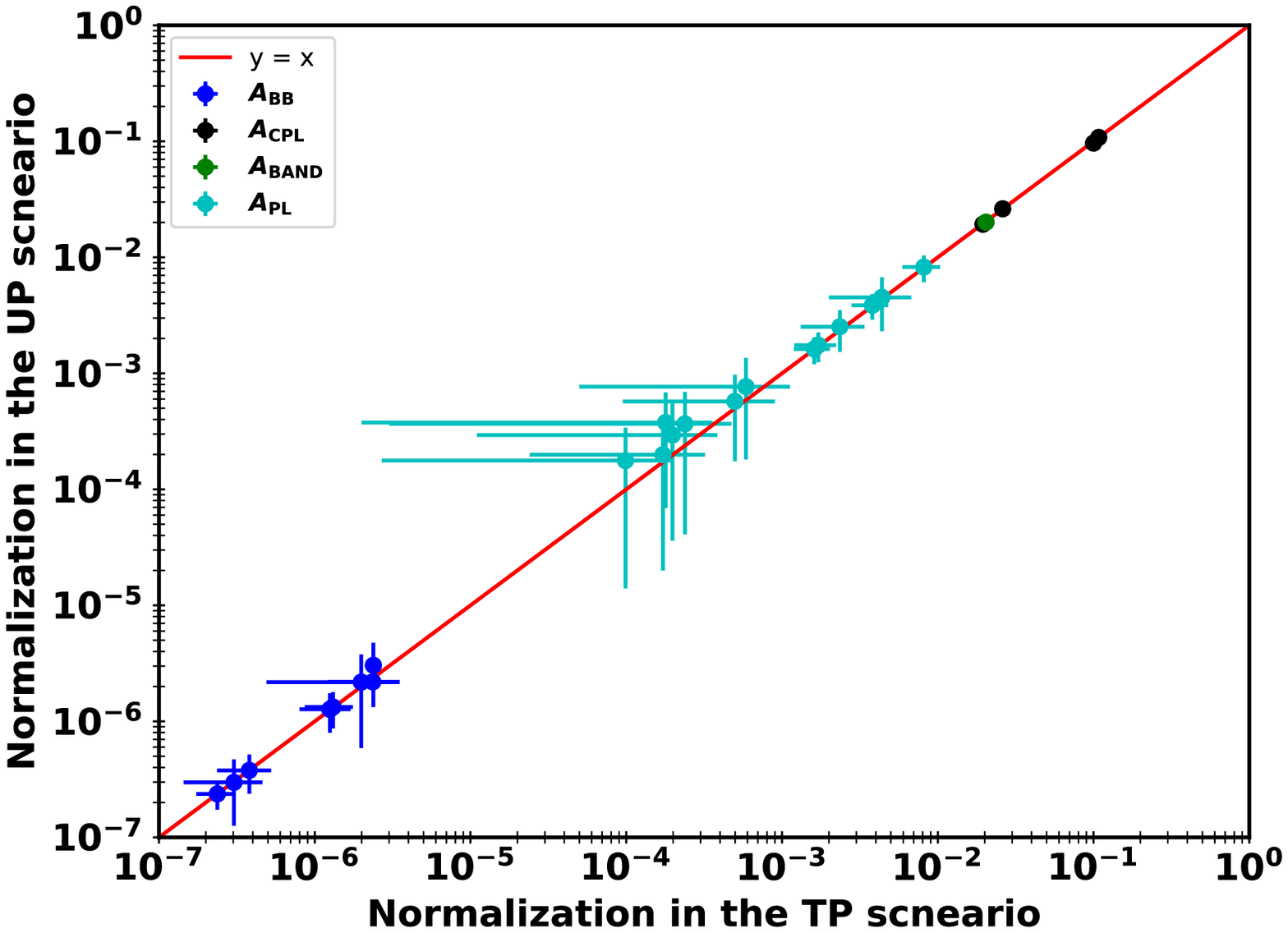}}
	\subfigure[]{
		\includegraphics[width=0.45\textwidth]{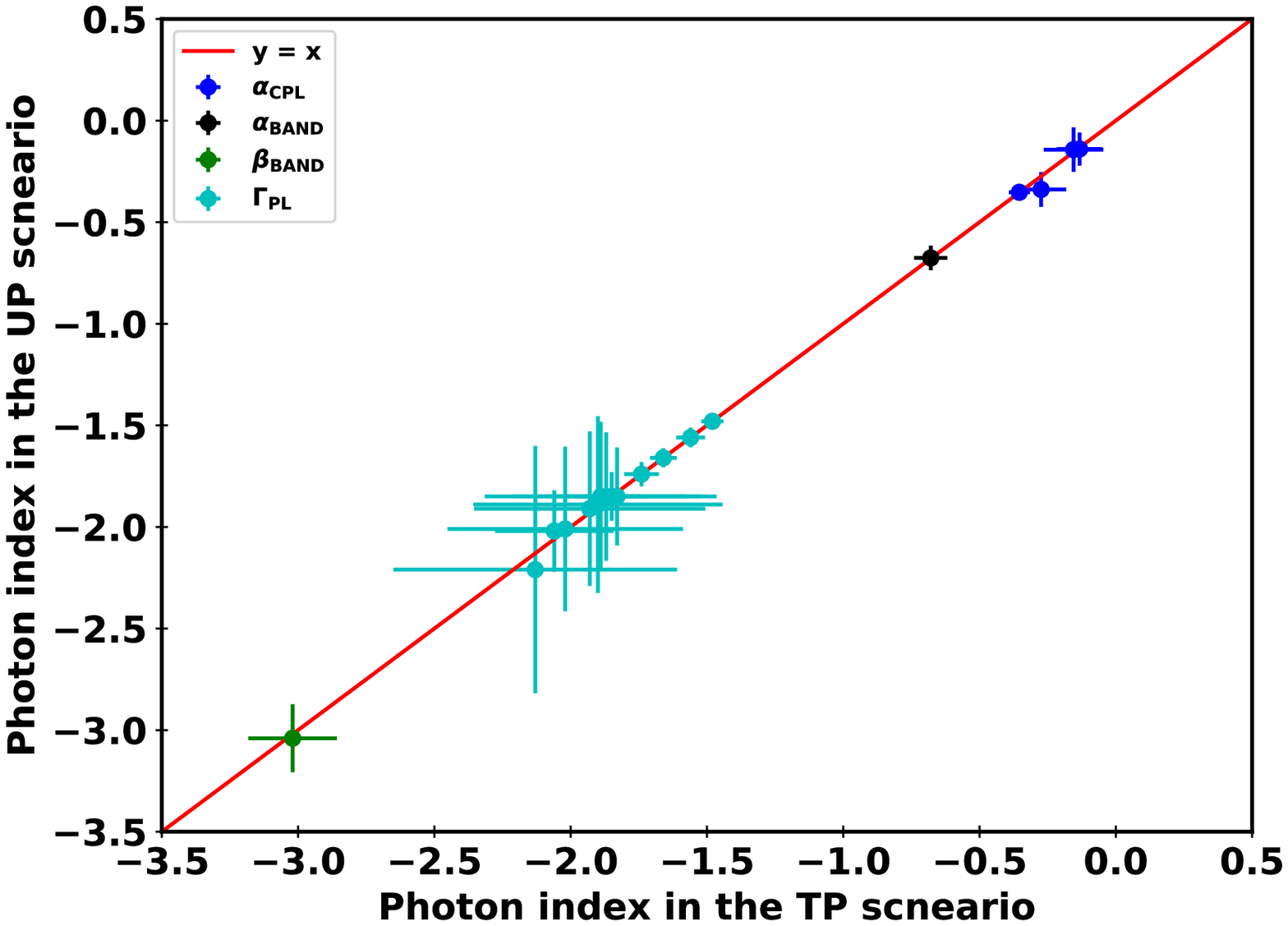}}
	\subfigure[]{
		\includegraphics[width=0.45\textwidth]{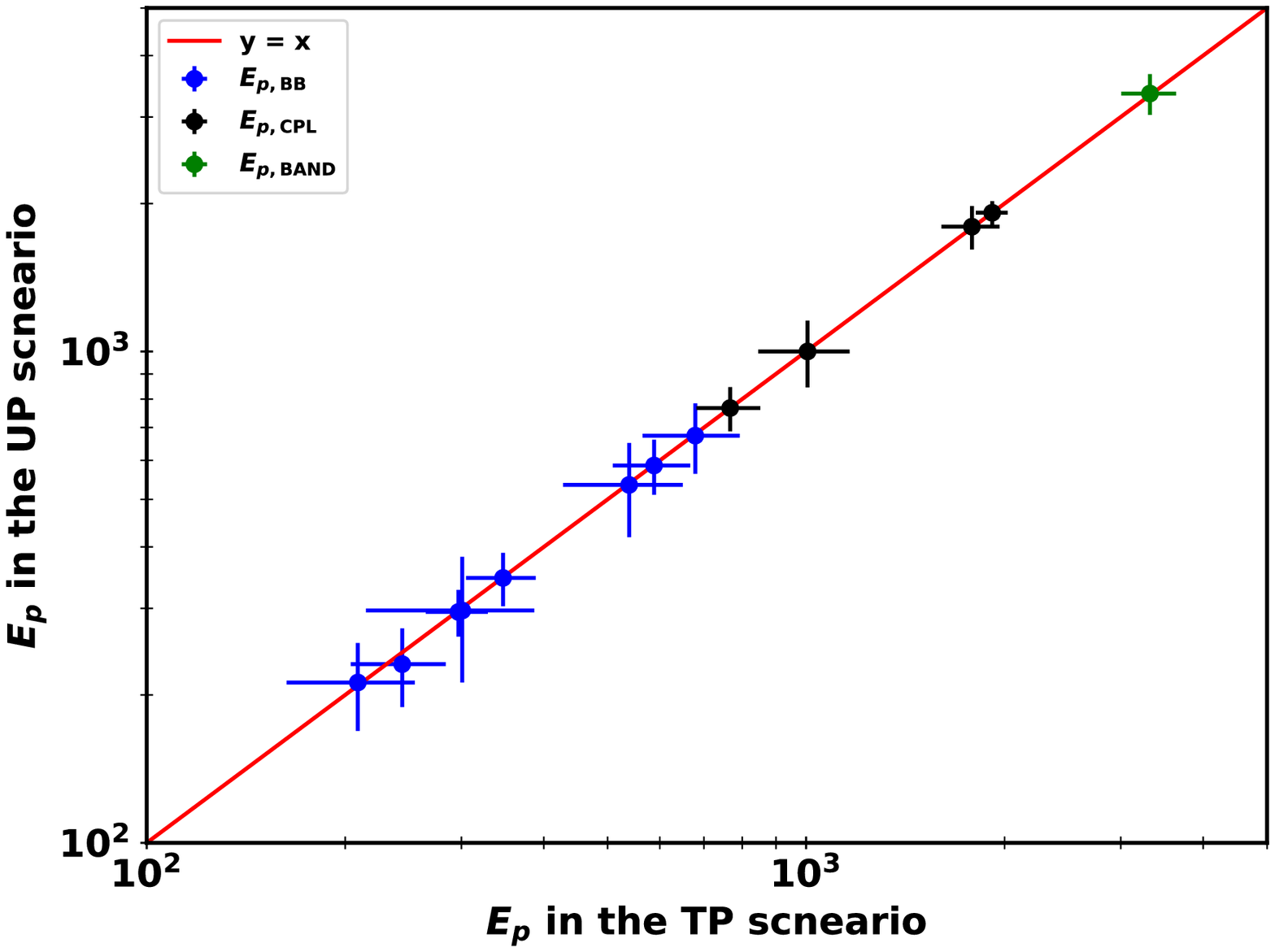}}
	\caption{Comparative derived parameters between the TP and UP scenario. The red solid line indicates $y = x$.}
\label{fig:compareUP}
\end{figure}

After all parameters available in both the UP scenario and the TP scenario, we thus plotted the correlations of the same parameters in two scenarios, which are shown in Figure \ref{fig:compareUP}. For parameters of the photon indices, the peak energies and the normalizations in both scenarios, they are almost lying at the equaling line ($y = x$).

In summary, the results of the Best 2C models in the UP scenario are consistent with that of the Best models in the TP scenario. Therefore, the results in the TP scenario are presented only in the main text.

\begin{deluxetable*}{c|cccccc|ccc}
\tablecaption{$\Delta{\rm BIC}$ and the best-fitting models in the UP scenario}
\label{table:comparisonUP}
\tablehead{
\colhead{GRB}
&\colhead{BB}
&\colhead{Band}
&\colhead{CPL}
&\colhead{BB+PL}
&\colhead{BAND+PL}
&\colhead{CPL+PL}
&\colhead{Best Model\tablenotemark{a}}
&\colhead{Best 1C Model\tablenotemark{b}}
&\colhead{Best 2C Model\tablenotemark{c}}
}
\startdata
081024B	&	$>$10\tablenotemark{d}	&	0	&	7	&	5	&	$>$10	&	7	&	BAND	&	BAND	&	BB+PL	\\
081102B	&	$>$10	&	6	&	0	&	6	&	$>$10	&	$>$10	&	CPL	&	CPL	&	BB+PL	\\
090227B	&	$>$10	&	$>$10	&	$>$10	&	$>$10	&	7	&	0	&	CPL+PL	&	--	&	CPL+PL	\\
090228A	&	$>$10	&	7	&	2	&	$>$10	&	5	&	0	&	CPL+PL	&	CPL	&	CPL+PL	\\
090510	&	$>$10	&	6	&	$>$10	&	$>$10	&	0	&	$>$10	&	BAND+PL	&	BAND	&	BAND+PL	\\
110728A	&	0	&	$>$10	&	6	&	8	&	$>$10	&	$>$10	&	BB	&	BB	&	BB+PL	\\
120830A	&	$>$10	&	6	&	0	&	$>$10	&	$>$10	&	8	&	CPL	&	CPL	&	CPL+PL	\\
120915A	&	0	&	$>$10	&	4	&	9	&	$>$10	&	$>$10	&	BB	&	BB	&	BB+PL	\\
140402A	&	0	&	$>$10	&	6	&	7	&	$>$10	&	$>$10	&	BB	&	BB	&	BB+PL	\\
141113A	&	0	&	7	&	1	&	9	&	$>$10	&	$>$10	&	BB	&	BB	&	BB+PL	\\
160709A	&	$>$10	&	0	&	$>$10	&	$>$10	&	2	&	1	&	BAND	&	BAND	&	CPL+PL	\\
171011C	&	0	&	$>$10	&	4	&	8	&	$>$10	&	$>$10	&	BB	&	BB	&	BB+PL	\\
190515A	&	1	&	6	&	0	&	9	&	$>$10	&	$>$10	&	CPL	&	CPL	&	BB+PL	\\
\enddata
\tablenotetext{a}{Best-fitting model with $\Delta{\rm BIC} = 0$}
\tablenotetext{b}{Best-fitting model with the lowest $\Delta{\rm BIC}$ among models of BB, BAND and CPL}
\tablenotetext{c}{Best-fitting model with the lowest $\Delta{\rm BIC}$ among models of BB+PL, BAND+PL and CPL+PL}
\tablenotetext{d}{$>$10 represents the best model against this candidate model}
\end{deluxetable*}

\begin{deluxetable*}{c|ccc|cc}
\label{table:bestfitUP}
\tablecaption{Derived parameter values of the Best 2C model in the UP scenario}
\tablehead{
\colhead{Class}
&\colhead{Main component}
&\colhead{}
&\colhead{}
&\colhead{Extra PL component}
&\colhead{}
}
\startdata
BB+PL	&	$A_{\rm BB}$\tablenotemark{a}	&	  &	$E_{\rm p,BB}$\tablenotemark{b}	&	 $A_{\rm PL}$\tablenotemark{c}	&	 $\Gamma_{\rm PL}$	\\
081024B	&	21.8	$\pm$	15.9	&		--		&	297	$\pm$	85	&	175.0	$\pm$	50.1	&	-1.74	$\pm$	 0.06	 \\
081102B	&	13.3	$\pm$	4.6	&		--		&	295	$\pm$	32	&	162.0	$\pm$	42.3	&	-1.85	$\pm$	0.12	 \\
110728A	&	21.9	$\pm$	16.8	&		--		&	231	$\pm$	42	&	37.7	$\pm$	30.8	&	-1.91	$\pm$	 0.38	 \\
120915A	&	12.7	$\pm$	4.7	&		--		&	346	$\pm$	43	&	17.7	$\pm$	16.3	&	-1.85	$\pm$	0.37	 \\
140402A	&	3.8	$\pm$	1.4	&		--		&	586	$\pm$	75	&	29.4	$\pm$	25.8	&	-2.21	$\pm$	0.61	 \\
141113A	&	3.0	$\pm$	1.7	&		--		&	535	$\pm$	116	&	76.8	$\pm$	58.7	&	-1.85	$\pm$	0.32	 \\
171011C	&	30.4	$\pm$	17.1	&		--		&	212	$\pm$	43	&	19.9	$\pm$	17.9	&	-1.89	$\pm$	 0.43	 \\
190515A	&	2.4	$\pm$	0.6	&		--		&	674	$\pm$	110	&	57.3	$\pm$	39.8	&	-1.85	$\pm$	0.24	 \\
\hline
CPL+PL	&	$A_{\rm CPL}$\tablenotemark{a}	&	 $\alpha_{\rm CPL}$ &	$E_{\rm p,CPL}$\tablenotemark{b}	&	 $A_{\rm PL}$\tablenotemark{c}	&	$\Gamma_{\rm PL}$	\\
090227B	&	10.8	$\pm$	0.5	&	-0.35	$\pm$	0.04	&	1915	$\pm$	106	&	824.0	$\pm$	214.0	&	-1.48	 $\pm$	0.04	\\
090228A	&	9.6	$\pm$	0.4	&	-0.34	$\pm$	0.09	&	767	$\pm$	79	&	452.0	$\pm$	221.0	&	-2.02	$\pm$	 0.20	\\
120830A	&	1.9	$\pm$	0.1	&	-0.14	$\pm$	0.11	&	1000	$\pm$	155	&	36.7	$\pm$	32.6	&	-2.01	 $\pm$	 0.40	\\
160709A	&	2.6	$\pm$	0.2	&	-0.14	$\pm$	0.08	&	1794	$\pm$	182	&	386.0	$\pm$	95.5	&	-1.66	 $\pm$	 0.05	\\
\hline
BAND+PL	&	$A_{\rm BAND}$\tablenotemark{a}	&	 $\alpha_{\rm BAND}$  &	$E_{\rm p,BAND}$\tablenotemark{b}		 &	 $A_{\rm PL}$\tablenotemark{c}	&	$\Gamma_{\rm PL}$	\\
&		&	($\beta_{\rm BAND}$)	&		&		&		\\
090510	&	2.0	$\pm$	0.2	&	-0.68	$\pm$	0.06	&	3348	$\pm$	318	&	252.0	$\pm$	98.9	&	-1.56	 $\pm$	 0.05	\\
&		&	(-3.04	$\pm$	0.17)	&		&		&		\\
\enddata
\tablenotetext{a}{Normalization for the main components, $A_{\rm BB}$ in unit of $10^{-7} {\rm ph\ keV^{-1} cm^{-2}\ s^{-1}}$, $A_{\rm CPL}$ and $A_{\rm BAND}$ in unit of $10^{-2} {\rm ph\ keV^{-1} cm^{-2}\ s^{-1}}$}
\tablenotetext{b}{Peak energy of the $E^2 dN/dE$ spectrum in unit of {\rm keV}}
\tablenotetext{c}{Normalization for the extra components in unit of $10^{-5} {\rm ph\ keV^{-1} cm^{-2}\ s^{-1}}$}
\end{deluxetable*}

\clearpage

\section{Spectral energy distributions with the best-fitting model for the other 10 SGRBs}
\label{appendixSED}

\begin{figure}[p]
\label{fig:bestfitleft1}
	\centering
	\subfigure[]{
		\includegraphics[width=0.45\textwidth]{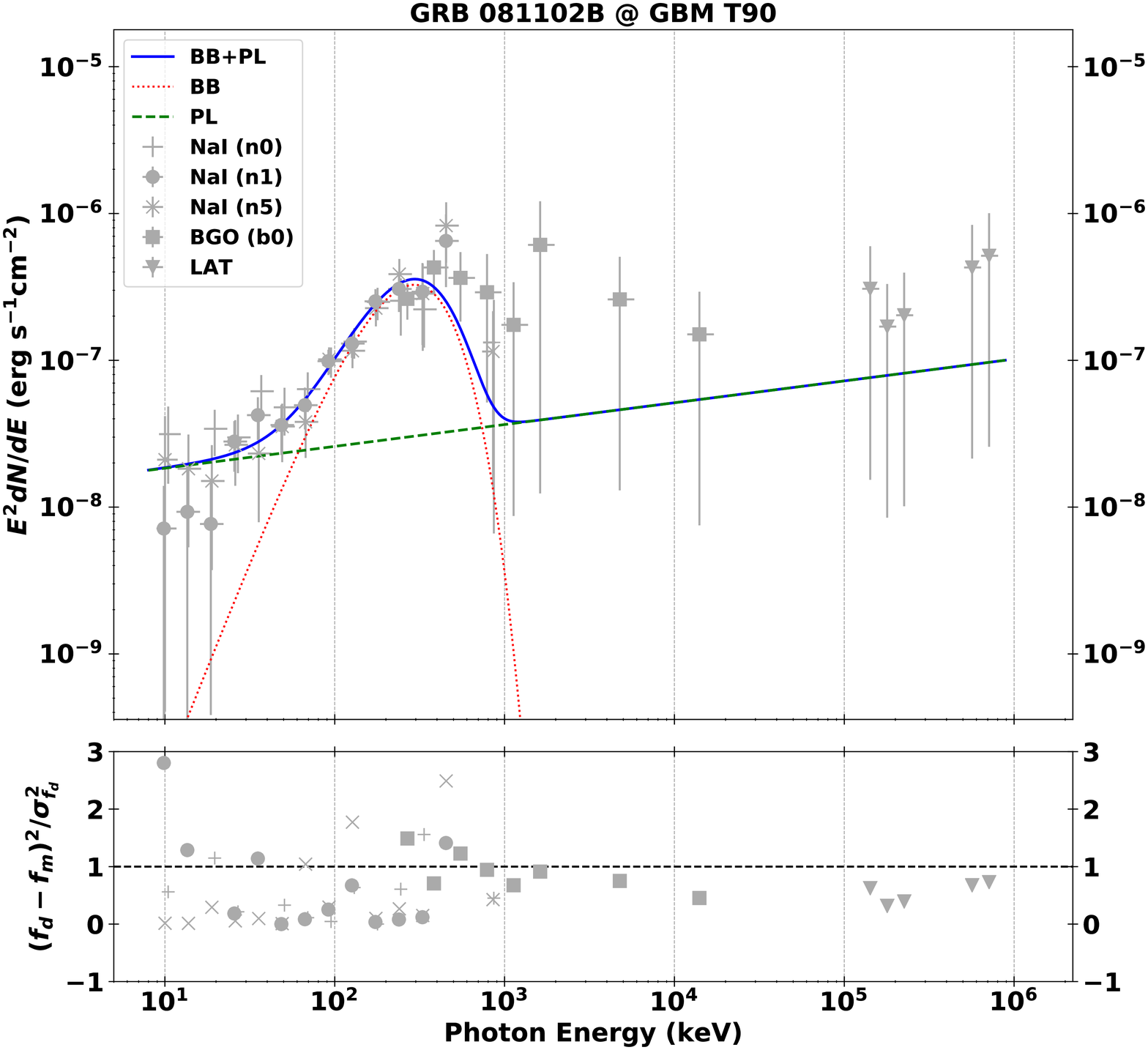}}
	\subfigure[]{
		\includegraphics[width=0.45\textwidth]{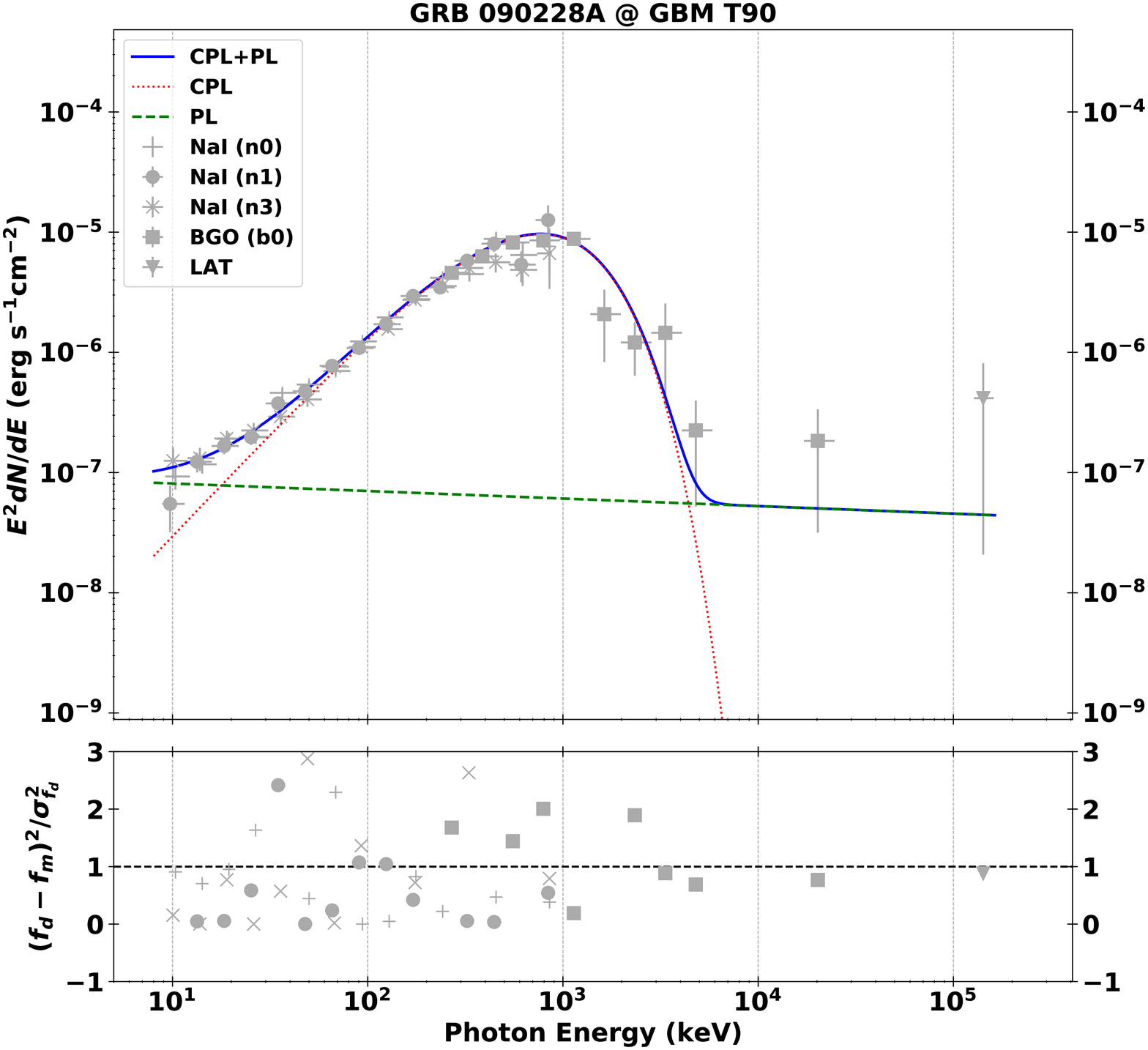}}
	\subfigure[]{
		\includegraphics[width=0.45\textwidth]{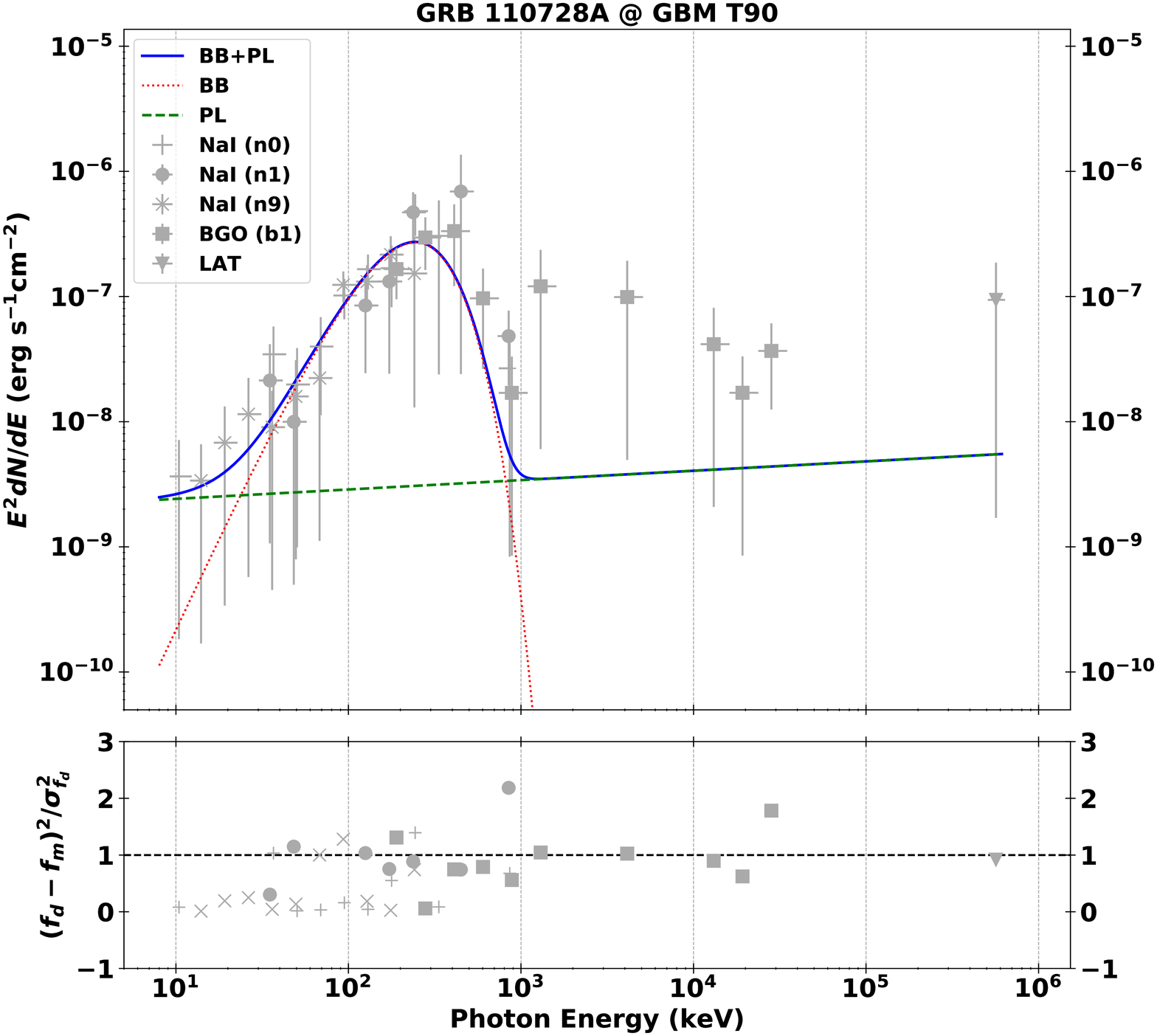}}
	\subfigure[]{
		\includegraphics[width=0.45\textwidth]{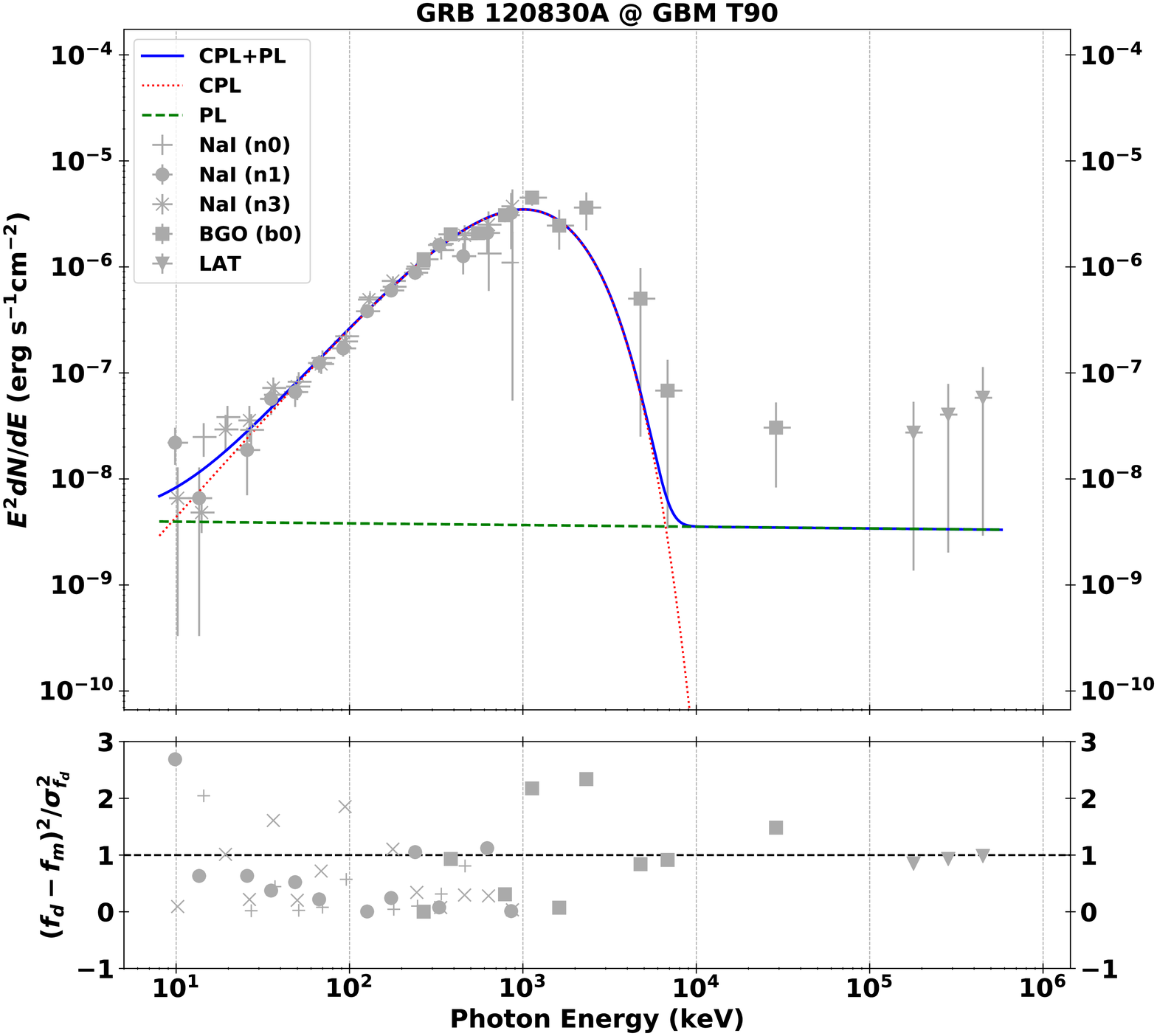}}
	\subfigure[]{
		\includegraphics[width=0.45\textwidth]{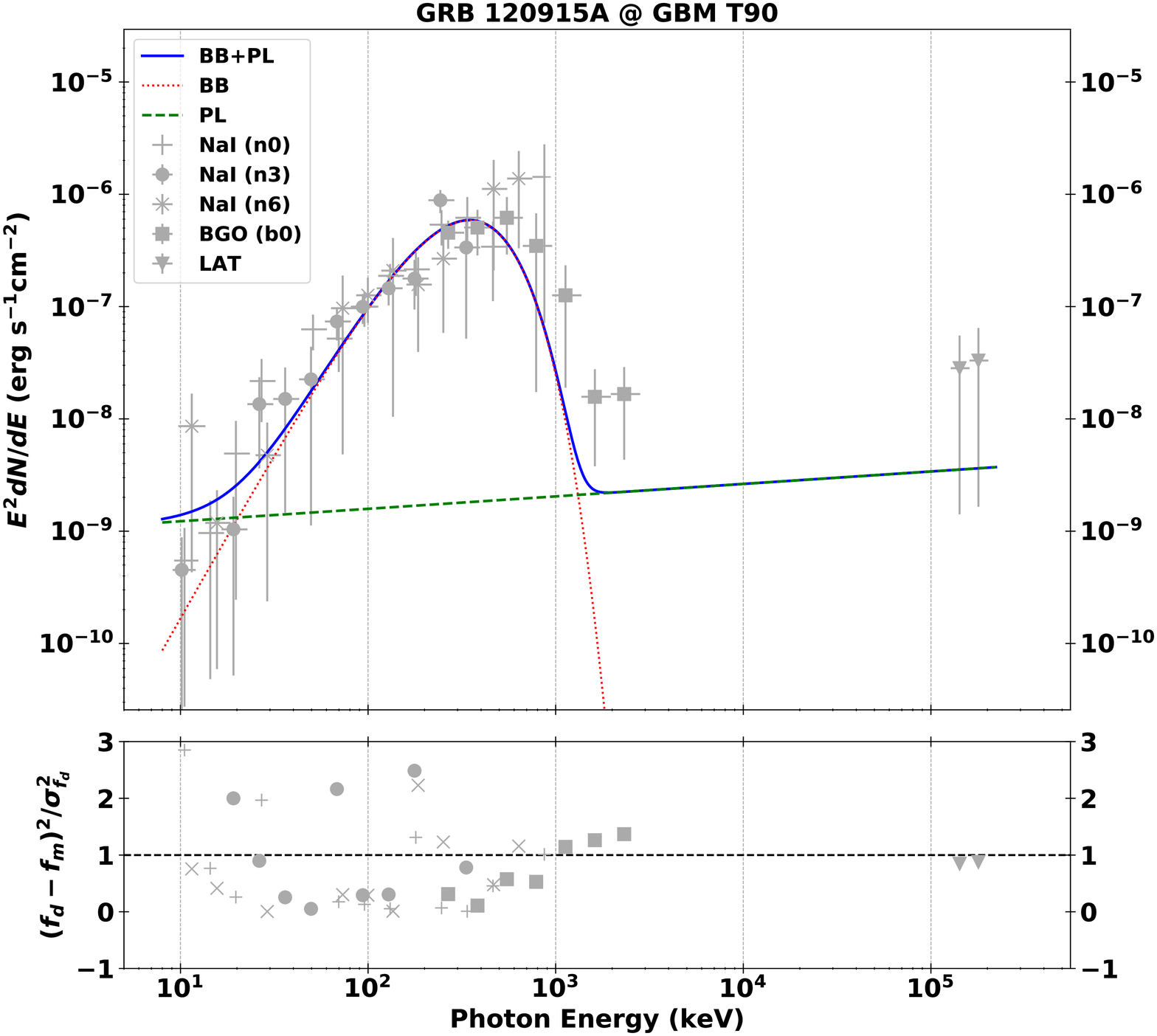}}
	\subfigure[]{
		\includegraphics[width=0.45\textwidth]{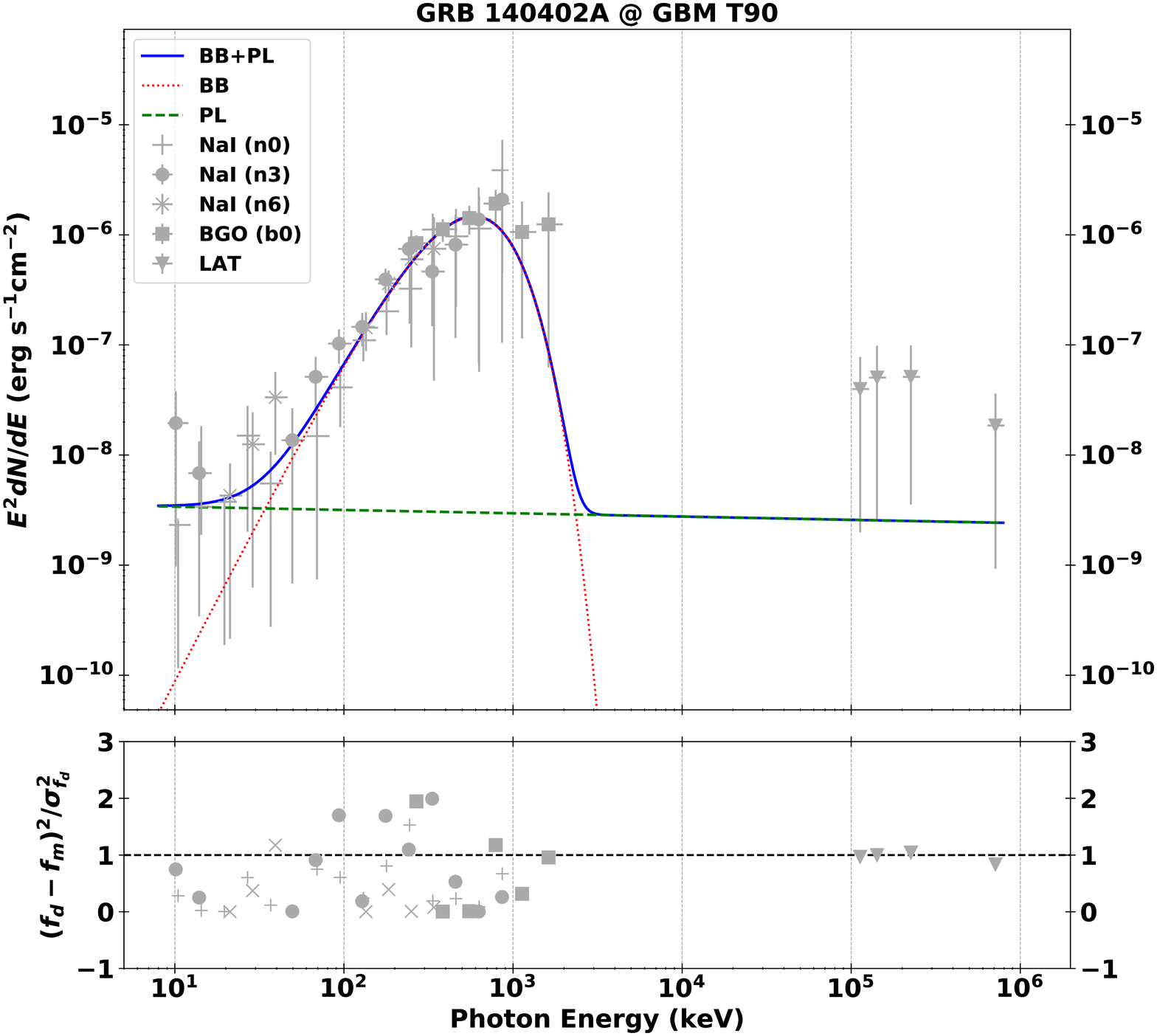}}
	\caption{Same as the SEDs in Figure \ref{fig:bestfit}, but for GRBs (a) 081102B, (b) 090228A, (c) 110728A, (d) 120830A, (e) 120915A and (f) 140402A.}
\end{figure}

\begin{figure}[p]
\label{fig:bestfitleft2}
	\centering
	\subfigure[]{
		\includegraphics[width=0.45\textwidth]{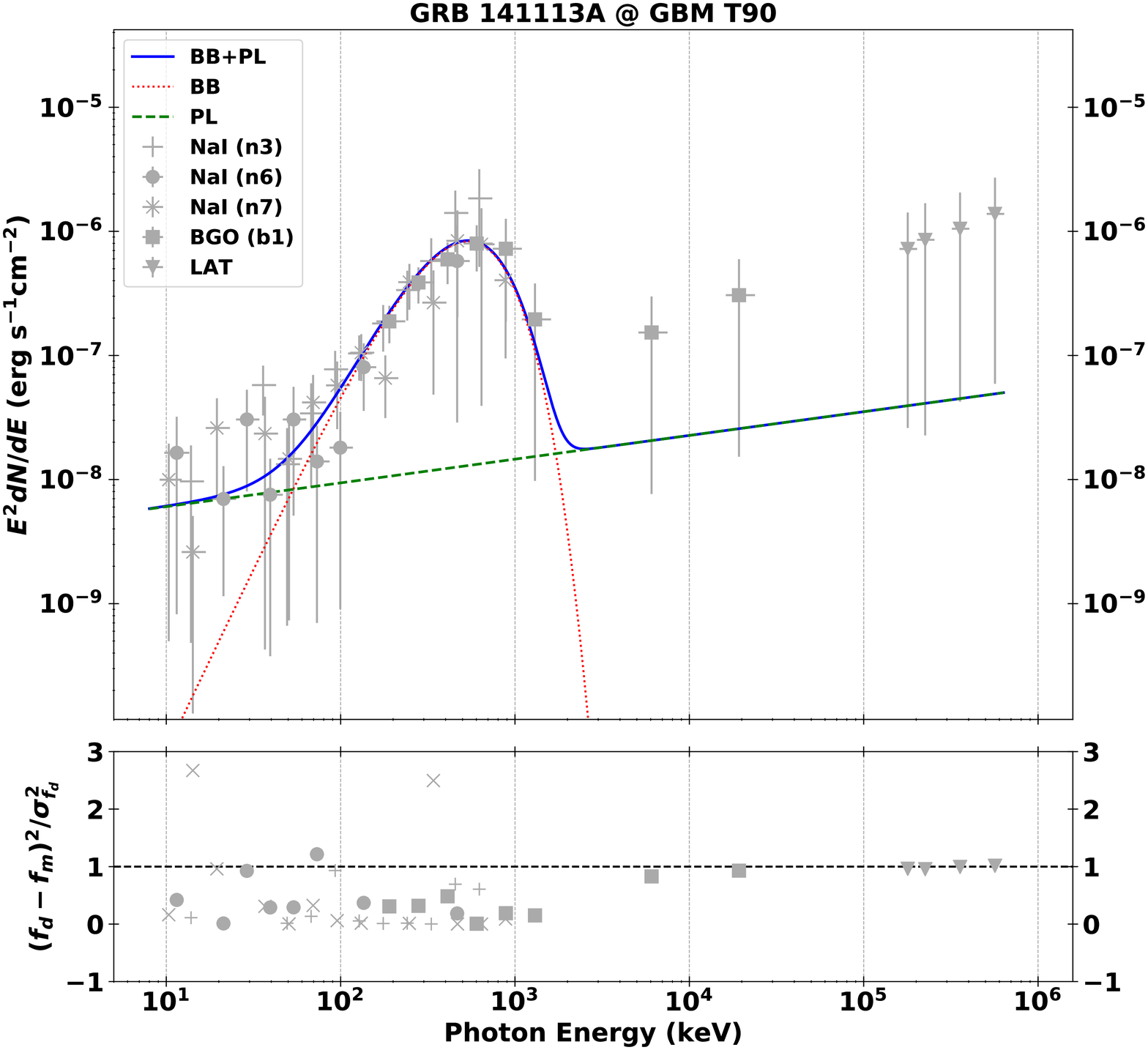}}
	\subfigure[]{
		\includegraphics[width=0.45\textwidth]{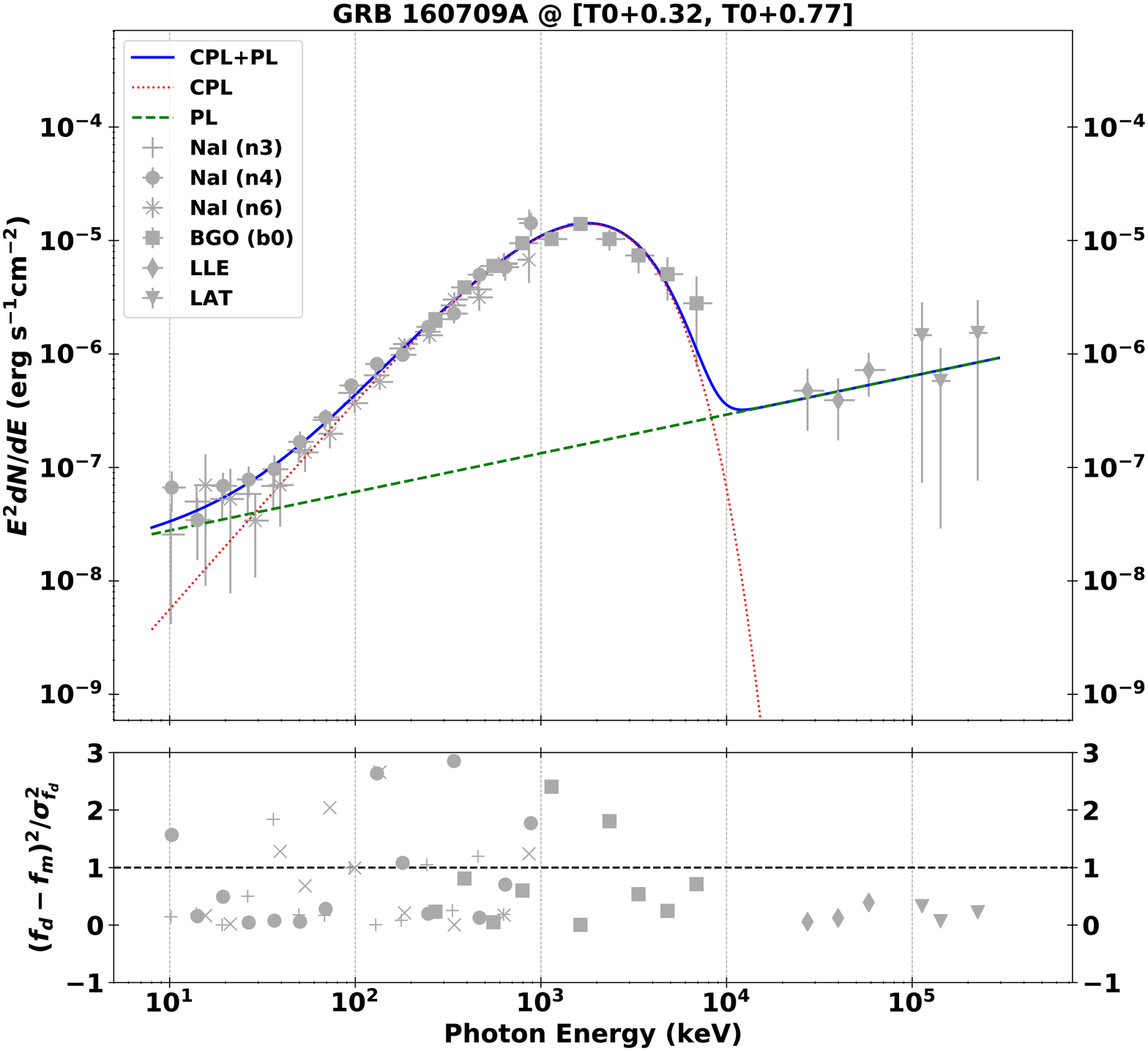}}
	\subfigure[]{
		\includegraphics[width=0.45\textwidth]{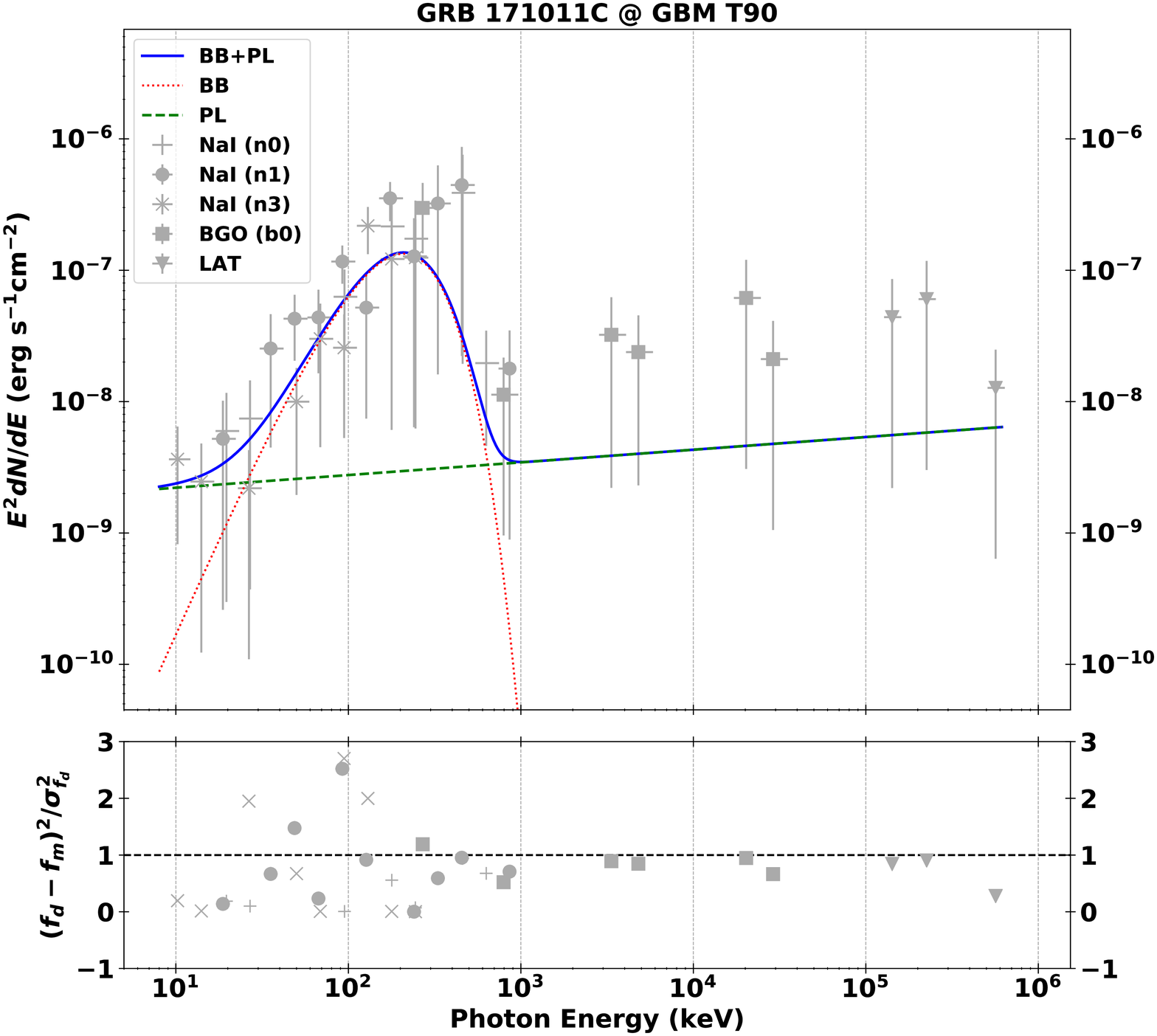}}
	\subfigure[]{
		\includegraphics[width=0.45\textwidth]{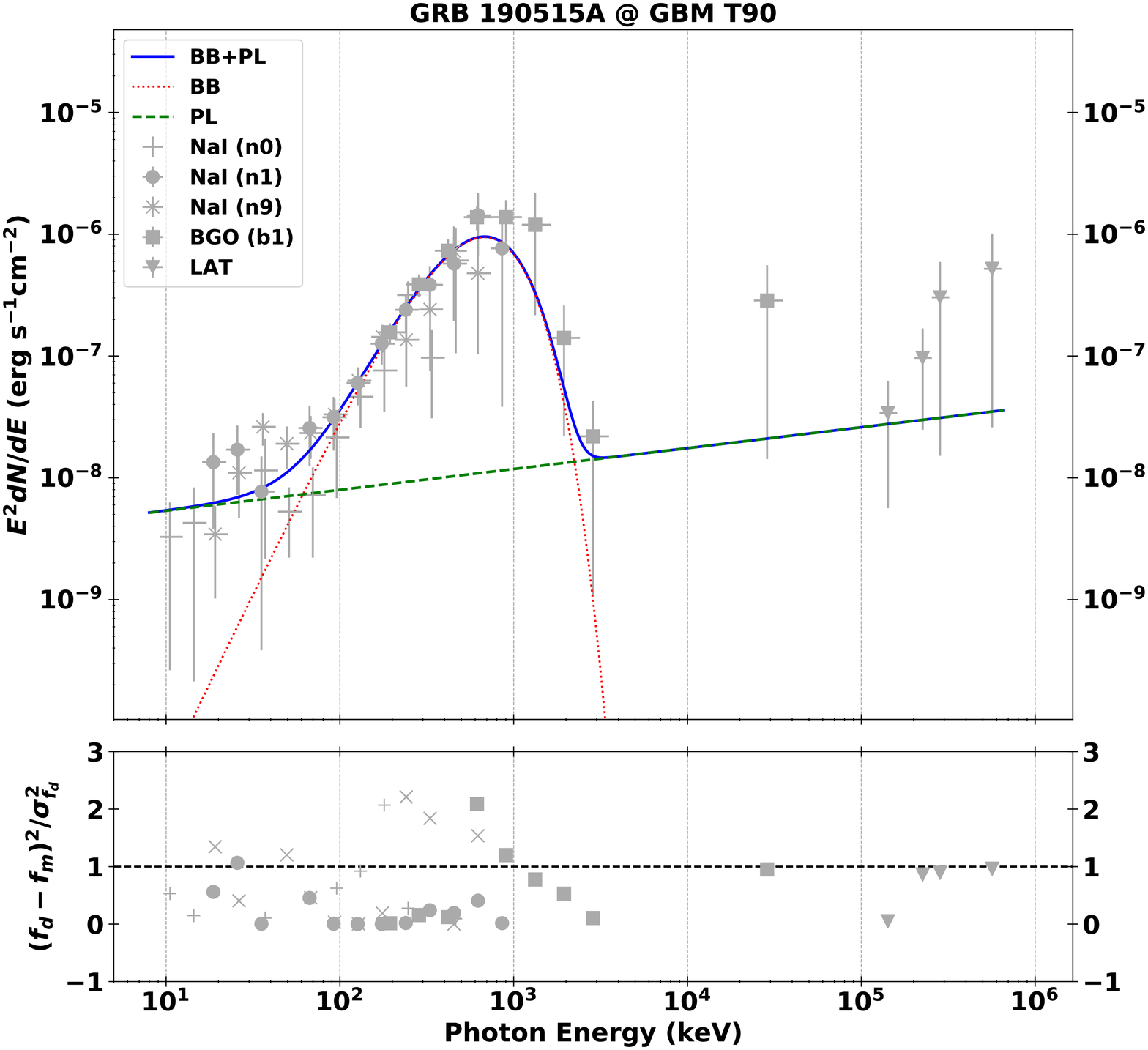}}
	\caption{Same as the SEDs in Figure \ref{fig:bestfit}, but for GRBs (a) 141113A, (b) 160709A, (c) 171011C and (d) 190515A.}
\end{figure}

\end{document}